\documentclass[twocolumn]{aastex63}

\usepackage{amsthm,latexsym,amssymb,amsmath, amsfonts}
\usepackage{xcolor}
\usepackage{subfloat}
\usepackage{graphicx}
\newcommand{\rutgers}{Rutgers University, Department of Physics and Astronomy, 136 Frelinghuysen Road, Piscataway, NJ 08854, USA}
\newcommand{\Drizzlepac}{\texttt{DrizzlePac 3.0}}
\newcommand{\TweakReg}{\texttt{TweakReg}}
\newcommand{\Astrodrizzle}{\texttt{AstroDrizzle}}
\newcommand{\Dolphot}{\texttt{DOLPHOT}}
\newcommand{\Beast}{\texttt{BEAST}}

\newcommand{\hO}{$H_0$}
\newcommand{\blueslope}{-1.15}
\newcommand{\redslope}{-2.16}
\newcommand{\bluezp}{-4.87}
\newcommand{\redzp}{-5.79}

\begin{document}

\title{An Empirical Calibration of the Tip of the Red Giant Branch Distance Method in the Near Infrared. I. HST WFC3/IR F110W and F160W Filters}

\author[0000-0002-8092-2077]{Max J. B. Newman}
\affiliation{\rutgers}

\author[0000-0001-5538-2614]{Kristen B. W. McQuinn}
\affiliation{\rutgers}

\author[0000-0003-0605-8732]{Evan D. Skillman}
\affiliation{University of Minnesota, Minnesota Institute for Astrophysics, School of Physics and Astronomy, 116 Church Street, S.E., Minneapolis,\\
MN 55455, USA}

\author[0000-0003-4850-9589]{Martha L. Boyer}
\affiliation{Space Telescope Science Institute, 3700 San Martin Drive, Baltimore, MD 21218, USA}

\author[0000-0002-2970-7435]{Roger E. Cohen}
\affiliation{\rutgers}

\author[0000-0001-8416-4093]{Andrew E. Dolphin}
\affiliation{Raytheon Technologies, 1151 E. Hermans Road, Tucson, AZ 85756, USA}
\affiliation{Steward Observatory, University of Arizona, 933 North Cherry Avenue, Tucson, AZ 85721, USA}

\newcommand{\princeton}{Department of Astrophysical Sciences, Princeton University, 4 Ivy Lane, Princeton, NJ 08544, USA}
\newcommand{\carnegie}{The Observatories of the Carnegie Institution for Science, 813 Santa Barbara Street, Pasadena, CA 91101, USA}

\author[0000-0003-4122-7749]{O. Grace Telford}
\altaffiliation{Carnegie-Princeton Fellow}
\affiliation{\princeton}
\affiliation{\carnegie}
\affiliation{\rutgers}

\correspondingauthor{Max J. B. Newman}
\email{mjn125@physics.rutgers.edu}

\begin{abstract}
The Tip of the Red Giant Branch (TRGB)-based distance method in the I band is one of the most efficient and precise techniques for measuring distances to nearby galaxies ($D\lesssim15$~Mpc). The TRGB in the near infrared (NIR) is 1 to 2~magnitudes brighter relative to the I band, and has the potential to expand the range over which distance measurements to nearby galaxies are feasible. Using Hubble Space Telescope (HST) imaging of 12 fields in 8 nearby galaxies, we determine color-based corrections and zero points of the TRGB in the Wide Field Camera 3 IR (WFC3/IR) F110W and F160W filters. First, we measure TRGB distances in the I band equivalent Advanced Camera System (ACS) F814W filter from resolved stellar populations with the HST. The TRGB in the ACS F814W filter is used for our distance anchor and to place the WFC3/IR magnitudes on an absolute scale. We then determine the color dependence (a proxy for metallicity/age) and zero point of the NIR TRGB from photometry of WFC3/IR fields which overlap with the ACS fields. The new calibration is accurate to $\sim1\%$ in distance, relative to the F814W TRGB. Validating the accuracy of the calibrations, we find that the distance modulus for each field using the NIR TRGB calibration agrees with the distance modulus of the same fields as determined from the F814W TRGB. This is a JWST preparatory program and the work done here will directly inform our approach to calibrating the TRGB in JWST NIRCam and NIRISS photometric filters.
\end{abstract}


\section{Tip of the Red Giant Branch Distances}
 Distances to galaxies are fundamental measurements and are essential to placing myriad galaxy properties (e.g., mass, luminosity, size, etc.) on absolute scales. In addition, extragalactic distance measurements are critical rungs in the cosmic distance ladder enabling local measurements of the Hubble Constant \citep[\hO; e.g.,][]{Freedman2019, Reid2019, Yuan2019, Freedman2021, Riess2022}. 
 
 Precise distances to nearby galaxies are measured with a variety of techniques using resolved stars including RR Lyrae variables, Horizontal Branch stars, Cepheids, and Tip of the Red Giant Branch (TRGB) stars \citep{Sandage2006, Pietrzynksi2008, Carretta2000, Beaton2018, Riess2022}. Among these methods, the TRGB-based distance method has emerged as one of the leading distance indicators to nearby galaxies as it offers a number of advantages over the other methods \citep[e.g.,][]{Mould1986, Freedman1988, DaCosta1990, Lee1993, Beaton2016, McQuinn2017}. It is one of the most efficient and precise methods, it can be employed in almost any galaxy type requiring only that they host a well-defined population of Red Giant Branch (RGB) stars, and the physical principles underlying the method are well understood \citep[e.g.,][]{Salaris1997, Madore1997, Salaris2002}.

The TRGB is defined as the end of the RGB stellar evolutionary phase. A low-mass star $\left(<2M_{\odot}\right)$ ceases burning hydrogen in its core as it exits the main-sequence and begins burning hydrogen in a shell around its inert, degenerate helium core. The star evolves up the RGB by expanding, increasing in luminosity, and dumping helium ash onto a still degenerate helium core. The core pressure and temperature build until a critical temperature of $\sim10^8$~K is reached, at which point helium fusion ignites, lifting the degeneracy. This point is defined as the TRGB. The sudden lifting of the degeneracy causes a release of energy referred to as the Helium Flash, after which the star moves onto either the Horizontal Branch phase, decreasing in brightness and increasing in surface temperature, or onto the Red Clump. Because the Helium Flash occurs at a specific stellar core mass, the TRGB luminosity is stable and predictable across stars with a wide range of properties and can be used as a standard candle
\citep[e.g.,][]{Lee1993, Beaton2018}.

While the bolometric luminosity of the TRGB is approximately constant, the apparent brightness of the TRGB in specific bandpasses will vary on a star-to-star basis due to differences in intrinsic stellar properties such as chemical content (i.e., metallicity), and, to a lesser extent, age. The impact of these variations in the apparent brightness of stars at the TRGB correlates with color and must be carefully quantified in order for the TRGB to be used as a precision distance indicator. 

 TRGB distances are typically measured in the I band where the brightness of the TRGB has been found to be approximately constant and metallicity/age-based corrections are modest, albeit with somewhat larger metallicity/age dependence for more metal-rich stellar populations \citep[e.g.,][]{Lee1993, Madore1995, Sakai1996}. This metallicity/age dependence is well understood and is accounted for in a number of TRGB calibrations using color-based corrections \citep[e.g.,][]{Mendez2002, Makarov2006, Rizzi2007, Jang2017, Freedman2021}.  
 
The TRGB is identified as a sharp discontinuity in the stellar luminosity function (LF), something that can be measured more accurately if the color correction is applied first. 
Two methods are commonly used to identify the location of the TRGB: (1) the Sobel edge-detection filter based on a zero-sum Sobel kernel, which is itself a discrete approximation of a gradient \citep[e.g.,][]{Lee1993, Madore1995, Sakai1996}, and (2) a maximum-likelihood (ML) fitter which assumes a simple power law with a cutoff in the TRGB region plus a power law of a second slope for a stellar population brighter than the TRGB \citep[e.g., asymptotic giant branch (AGB) stars; e.g.,][]{Mendez2002, Makarov2006}. Once the TRGB luminosity has been identified and measured, the distance to the galaxy can be determined using a zero point calibrated from previous studies \citep[e.g.,][]{Lee1993,Rizzi2007, Jang2017}.

As one moves from the I band to longer wavelengths in the near-infrared (NIR), the TRGB increases in brightness by $\sim$1--2~mag offering observational gains. If the NIR TRGB can be robustly calibrated, the TRGB-based distance method has the potential to be applied to galaxies at larger distances and over a greater volume. In addition, shorter observation times will be required to measure high-precision distances, further increasing the efficiency of the method. A TRGB calibration in the NIR is timely given the new and upcoming space-based telescopes designed to operate in the NIR, i.e., the James Webb Space Telescope (JWST) and the Nancy Grace Roman Observatory. 

In contrast to the TRGB in the I band, the luminosity of the NIR TRGB exhibits a greater metallicity/age dependence in the sense that the TRGB stars have a larger range in brightness as a function of metallicity/age \cite[e.g.,][]{Ferraro2000, Valenti2004, Salaris2005, McQuinn2019}. There have been previous calibrations of the color dependence of the NIR TRGB through a combination of theoretical stellar evolutionary models and photometric data \citep[e.g.,][]{Dalcanton2012a, Wu2014, Serenelli2017, Hoyt2018, Madore2018, Gorski2018, Durbin2020, Freedman2020, Madore2023b}. However, current theoretical models from stellar evolution libraries vary in their zero point and metallicity/age-dependent slopes of the NIR TRGB, and are not always self-consistent across different wavelengths \citep[e.g.,][]{Serenelli2017, Durbin2020}. 

In this work, we calibrate the NIR TRGB in the Hubble Space Telescope (HST) Wide Field Camera 3 (WFC3) F110W and F160W filters. Our calibration is entirely empirical as we do not use theoretical stellar evolution libraries, and is tied to the well understood F814W TRGB \citep{Jang2017, Freedman2021}. In addition, this program is a JWST preparatory program and the work done here will directly inform our approach to calibrating the TRGB in several JWST Near Infrared Camera (NIRCam) and Near Infrared Imager and Slitless Spectrograph (NIRISS) photometric filters (JWST-GO-1638; PI McQuinn). 

The paper is organized as follows: We present the HST observations, data reduction, photometry, artificial star tests (ASTs), and foreground-extinction corrected color magnitude diagrams (CMDs) in \S~\ref{sec:obs_and_phot}. We describe the TRGB-based distance method used to determine the distances to our sample, which anchor our calibration, and
briefly compare our distance measurements to TRGB distances in the literature in \S~\ref{sec:ACS_TRGB}. Next, we discuss our methods for calibrating the NIR TRGB color-based correction and zero point and we present the NIR TRGB calibration in the F110W and F160W filters in \S~\ref{sec:NIR_TRGB_Calibration}. Finally, we discuss our results in the context of previous calibrations and summarize our findings in \S~\ref{sec:discussion} and \S~\ref{sec:summ_conc}, respectively.

\section{ Observations and Data Reduction}\label{sec:obs_and_phot}
\subsection{Galaxy Sample}
The luminosity of the TRGB has a dependence on stellar metallicity/age, which changes as a function wavelength. In the F814W filter, the dependence is modest but is still important to consider not only in its calibration, but also in observing strategies. Specifically, the TRGB in F814W is relatively flat for metal poor stellar populations \citep[e.g., ][]{Beaton2016}, so the outer stellar fields of galaxies which are thought to be primarily metal poor are commonly targeted for TRGB distance measurements. Outer stellar fields offer additional advantages including (1) reduced impact from crowding effects where the apparent brightness of a star can be biased to higher or lower magnitudes in regions of relatively high stellar densities (e.g., in stellar disks), (2) there are fewer young stellar populations that can blur the TRGB, and (3) there is less internal extinction.

However, there is observational evidence that suggests galaxy halos can be populated with stars that span a wider range of metallicity. Recent studies of spiral halos have shown metallicity gradients  \citep[e.g.,][]{Ibata2014, Gilbert2014, Cohen2020}, high median metallicities (i.e., [Fe/H] $> -1.2$~dex) as far as 100~kpc from galaxy centers \citep[e.g.,][]{Conroy2019}, significant field-to-field variations in the mean metallicity of RGB stars for an individual galaxy (including along the major vs.\ minor axes), and metallicity variations of ~1 dex across a sample of 6 spiral halos \citep{Monachesi2016}.

In the NIR, because the luminosity of the TRGB has a more significant dependence on metallicity/age \citep[e.g.,][]{Salaris2005, Serenelli2017, McQuinn2019}, the impact of variations in metallicity/age on the TRGB luminosity must be carefully accounted for. We have crafted our calibration sample to explicitly account for these metallicity/age variations. Our calibration of the NIR TRGB is based on observations of galaxies with a wide range of  metallicities. The observations include new imaging of outer stellar fields in four nearby galaxies obtained for this program, and archival imaging of four low-mass galaxies. 
We supplemented the new observations with the archival data to increase the metallicity baseline of the sample. 

For these new observations, we made use of the separation of the HST Advanced Camera System/Wide Field Camera \citep[ACS/WFC; ][]{Ford1998} and WFC3/IR on the sky to image two outer stellar fields per galaxy with WFC3 in the parallel observing mode. This observing strategy enables us to account for potential changes in the stellar content of outer stellar fields. Field one was selected with ACS, and the roll angle was set such that WFC3/IR imaged another region of the outer stellar field. A second visit was planned for each field such that orientation of ACS and WFC3/IR were flipped. Figure~\ref{fig:original_3_color} presents GALEX images for each galaxy with footprint targets of ACS (green) and WFC3/IR (white). We placed tight roll angle constraints on our observations to ensure that each field was imaged with both instruments.

The archival data consist of HST observations of 4 nearby dwarf galaxies with overlapping images from the same set of filters as in our new observations. The archival data include imaging of both the stellar disk and part of the halos. Figure~\ref{fig:extension_3_color} shows GALEX images of the 4 dwarf galaxies with footprint targets of ACS (green) and WFC3/IR (white). The combined observations provide a uniform data set for our calibration. 

Table~\ref{tab:table1} presents a summary of the $HST$ observations including galaxy coordinates and observation details. The new observations were obtained as part of the JWST preparatory program HST-GO-15917. The archival observations were obtained over three different programs: ACS imaging is from  HST-GO-10210 and HST-GO-10915 \citep[ACS Nearby Galaxy Survey Treasury;][]{Dalcanton2009}, and the WFC3/IR imaging is part of  HST-GO-16162 (LUVIT; Boyer et al.\ in prep.).

\begin{deluxetable*}{lccccccccch}[!t]
    \tablecaption{Galaxy Sample and Summary of the $HST$ Observations}
    \tablewidth{0pt}
    \tabletypesize{\footnotesize}
    \setlength{\tabcolsep}{0.05in}
    \tablehead{
    Target & \colhead{R.A.} & \colhead{Decl.} & \colhead{A$_{V}$} & \multicolumn{2}{c}{ACS/WFC} & \colhead{PID} & \multicolumn{2}{c}{WFC3/IR} & \colhead{PID}\\
    & \colhead{(J2000)} & \colhead{(J2000)} & (mag) & \multicolumn{2}{c}{Exposure Time (s)} & & \multicolumn{2}{c}{Exposure Time (s)}\\
    & & & & \colhead{F606W} & \colhead{F814W}  & & \colhead{F110W}  & \colhead{F160W}
    }
    \startdata
    M81-1 (NGC3031-1) & 09:57:34.988 & +69:00:25.71 & 0.220 & $1167$  & $1167$  & HST-GO-15917 & $1498$  & $1448$  & HST-GO-15917 \\
    M81-2 (NGC3031-2) & 09:57:50.269 & +69 06 6.42 & 0.220 & $1167$  & $1167$  & HST-GO-15917 & $1498$  & $1448$  & HST-GO-15917 \\
    NGC253-1 & 00:46:16.545  & --25:27:54.12 & 0.051 & $1095$  & $918$  & HST-GO-15917 & $1348$  & $1298$  & HST-GO-15917 \\
    NGC253-2 & 00:46:40.877 & --25:30:27.46 & 0.051 &$1095$  & $918$  & HST-GO-15917 & $1348$  & $1298$  & HST-GO-15917 \\
    NGC300-1 & 00:54:07.000 & --37:33:19.61 & 0.035 & $1024$  & $1024$  & HST-GO-15917 & $1399$  & $1299$  & HST-GO-15917 \\
    NGC300-2 & 00:54:36.646 & --37:31:45.20 & 0.035 & $1024$  & $1024$  & HST-GO-15917 & $1399$  & $1299$  & HST-GO-15917 \\ 
    NGC2403-1 & 07:38:45.327 & +65:29:38.00 & 0.110 &$1167$  & $1167$  & HST-GO-15917 & $1498$  & $1448$  & HST-GO-15917 \\
    NGC2403-2 & 07:38:21.450 & +65:35 :12.32 & 0.110 &$1167$  & $1167$  & HST-GO-15917 & $1498$  & $1448$  & HST-GO-15917 \\
    Antlia Dwarf & 10:04:03.770 & -27:19:47.24 & 0.215 & $985$ & $1174$ & HST-GO-10210 & $1200$ & $1200$  & HST-GO-16162\\
    UGC8833 & 13:54:48.670 & +35:50:14.70 & 0.032 & $998$ & $1189$ & HST-GO-10210 & $2553$ & $2552$ & HST-GO-16162\\
    UGC9128 & 14:14:56.750 & +23.03:24.64 & 0.064 & $985$ & $1174$ & HST-GO-10210 & $2500$ & $2500$  & HST-GO-16162\\
    UGC9240 (DDO190) & 14:24:43.450& +44:31:40.52 & 0.034 & $2301$ & $2265$ & HST-GO-10915 & $2596$ & $2596$  & HST-GO-16162\\
    \enddata
    \tablecomments{Here the R.A. and Decl. are the coordinates of the ACS/WFC pointings. Foreground A$_{\text{V}}$ extinction values are from \cite{Schlafly2011}.}
    \label{tab:table1}
    \end{deluxetable*}

\begin{subfigures}
\begin{figure*}[!th]
	\centering
	\includegraphics[width=0.4\textwidth]{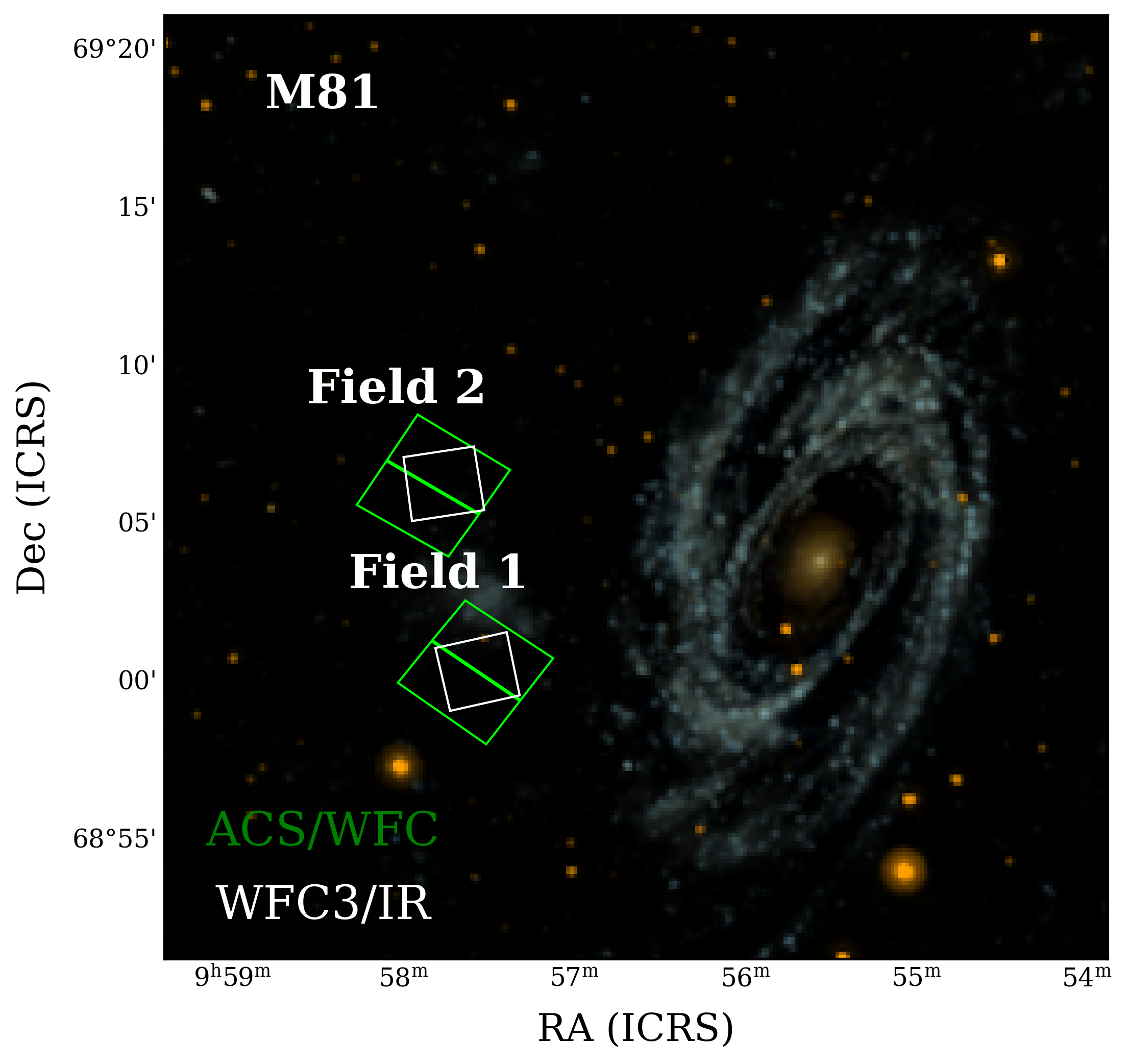}
    \includegraphics[width=0.4\textwidth]{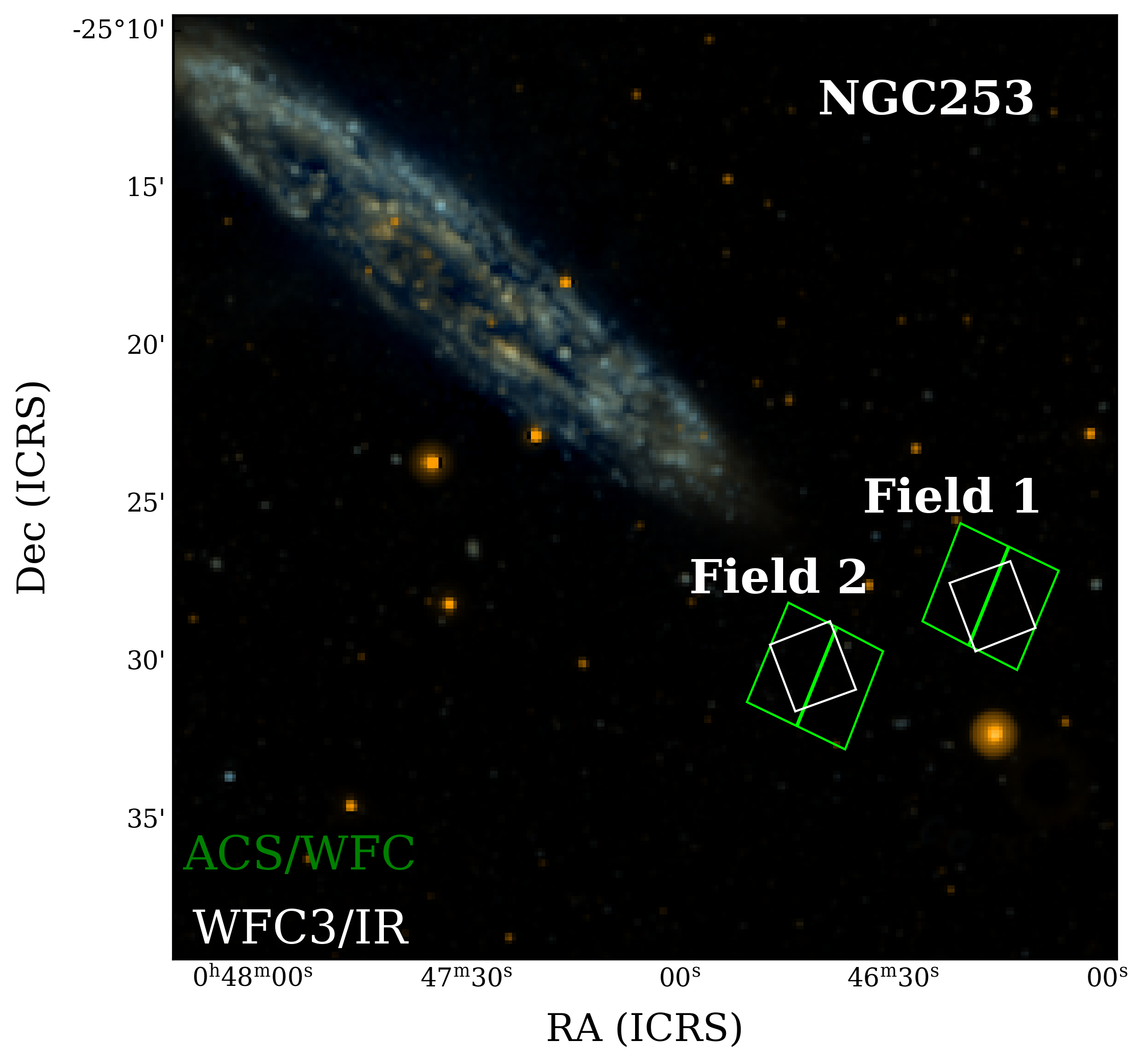}
	\includegraphics[width=0.4\textwidth]{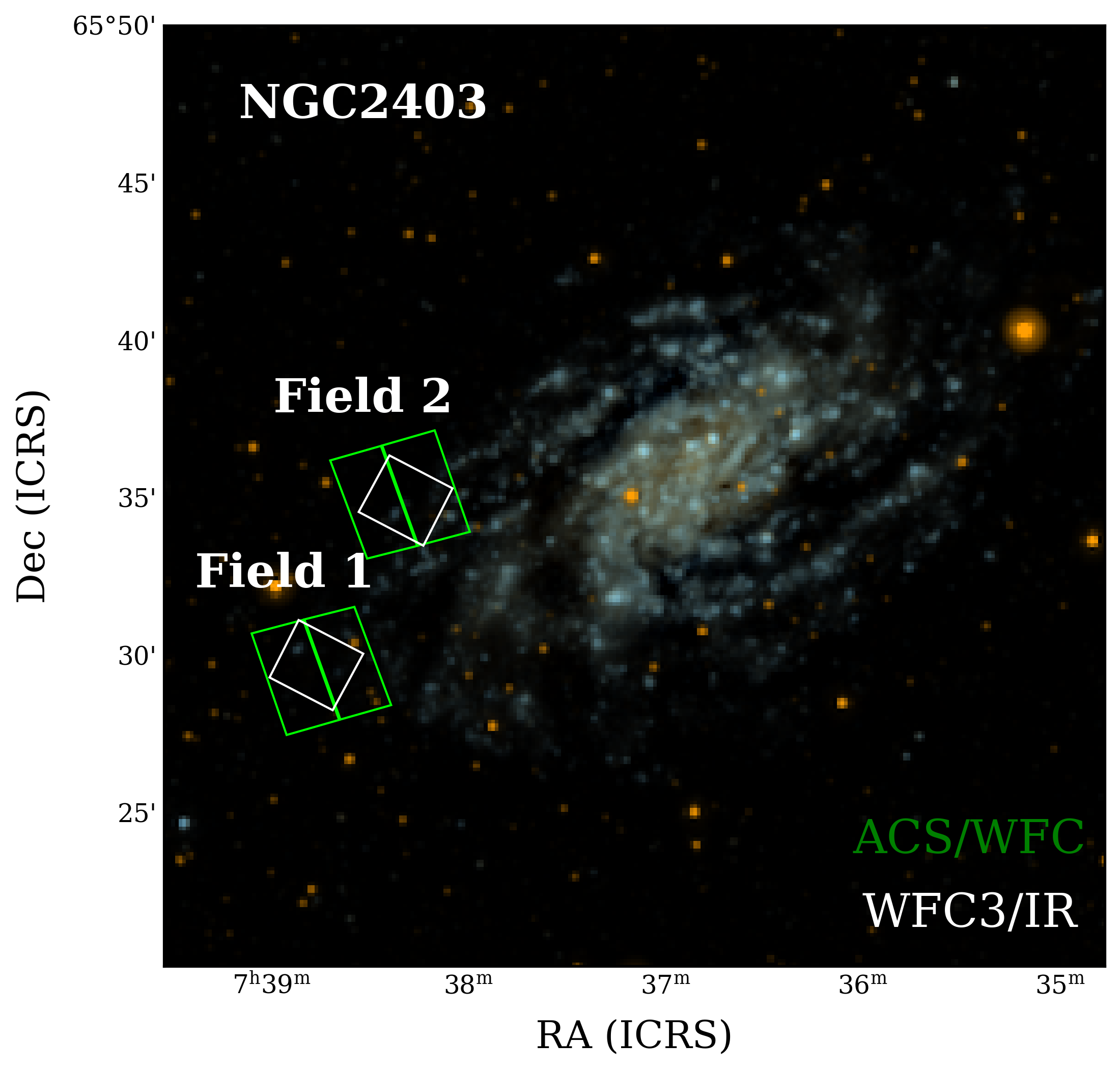}
    \includegraphics[width=0.4\textwidth]{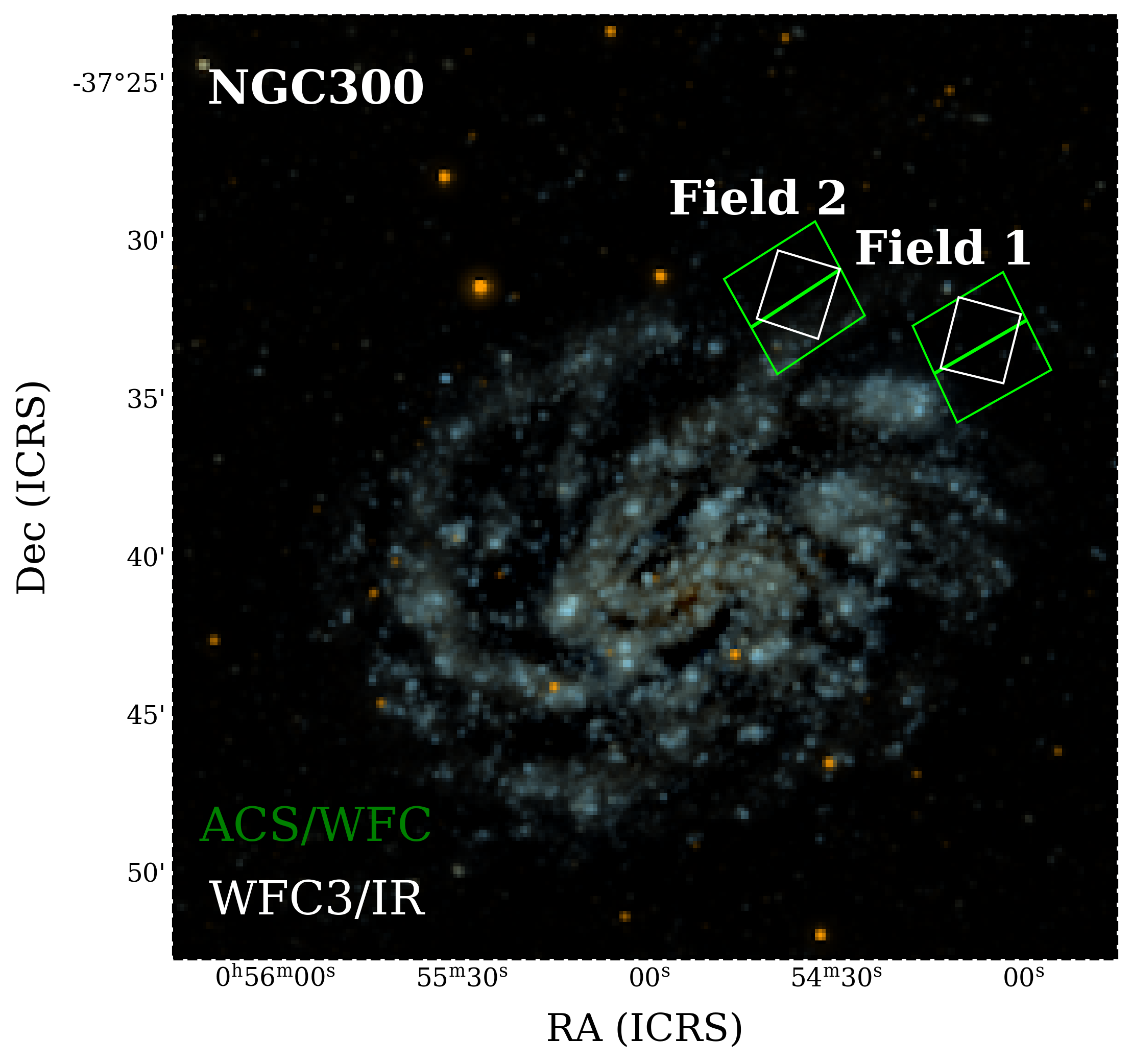}
	\caption{Clockwise from the top left: M81, NGC253, NGC300, NGC2403. Shown are color images using GALEX near-ultraviolet (NUV; blue) and far-ultraviolet (FUV; red) filters, and an average of the NUV and FUV filters (green), with the HST ACS and WFC3/IR instrument footprints indicated in green and white, respectively. Images are north up, east left. GALEX images were obtained courtesy of NASA/JPL-Caltech.}
	\label{fig:original_3_color}
\end{figure*}

\begin{figure*}[!th]
	\centering
	\includegraphics[width=0.4\textwidth]{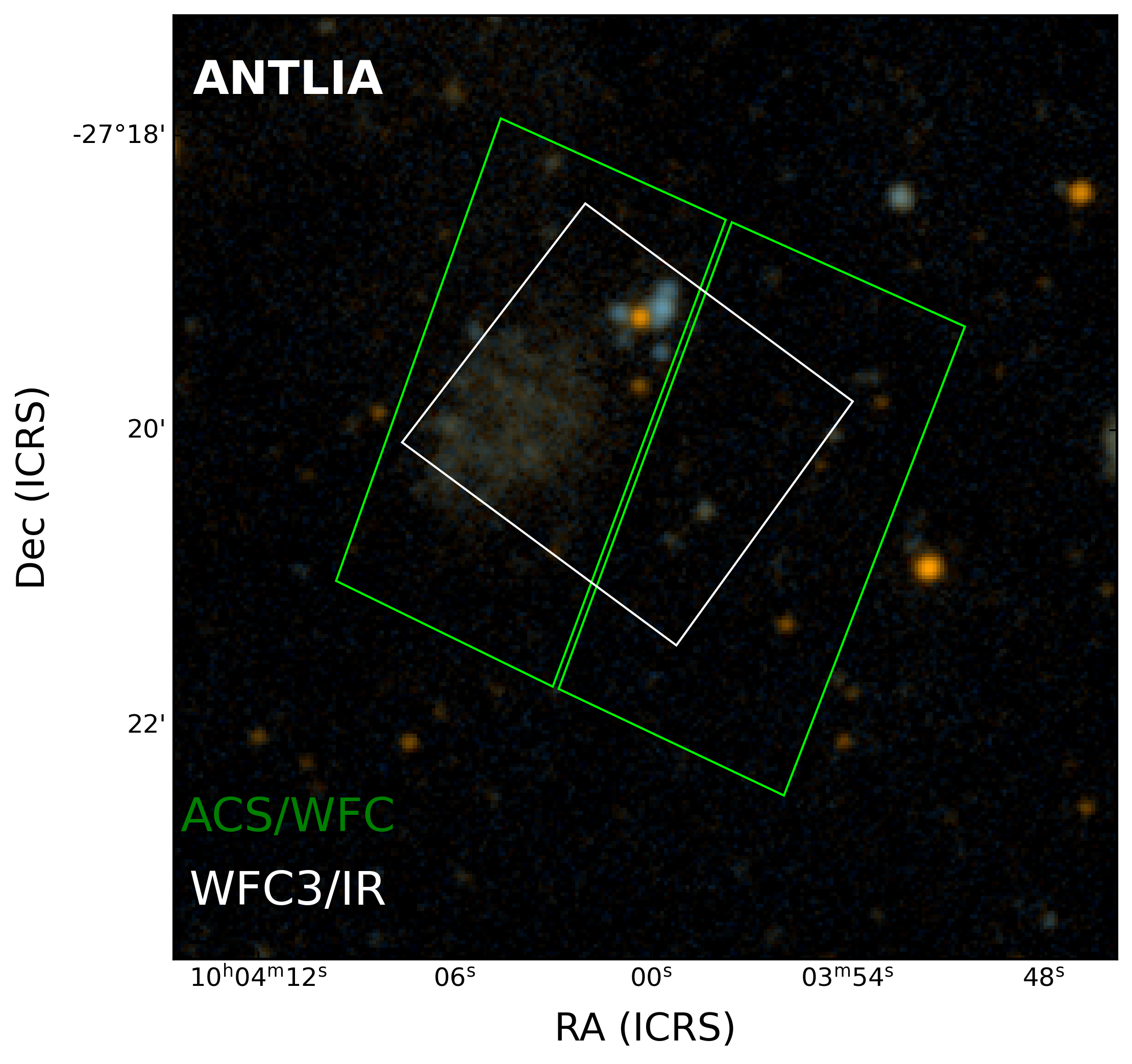}
    \includegraphics[width=0.4\textwidth]{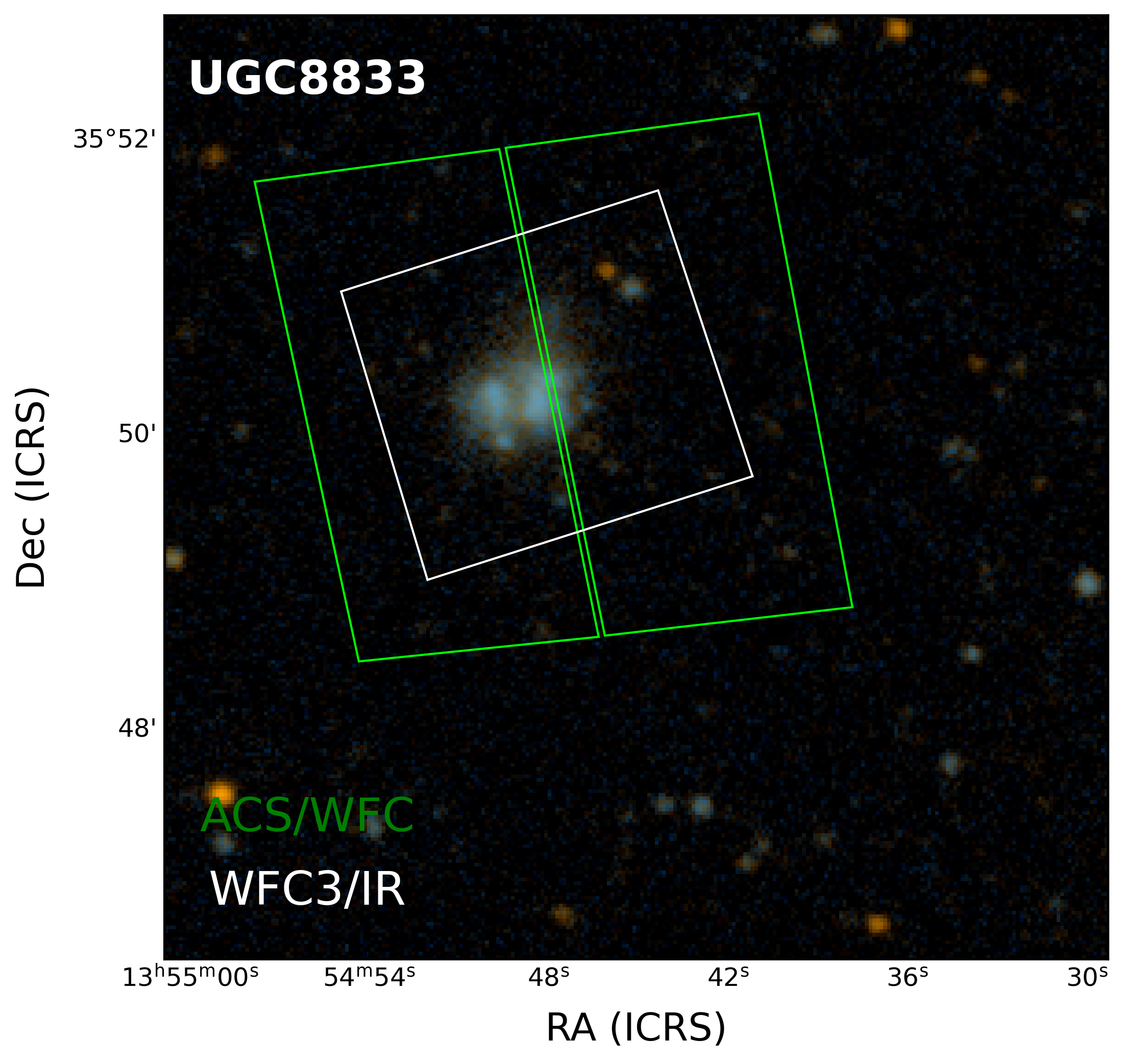}
	\includegraphics[width=0.4\textwidth]{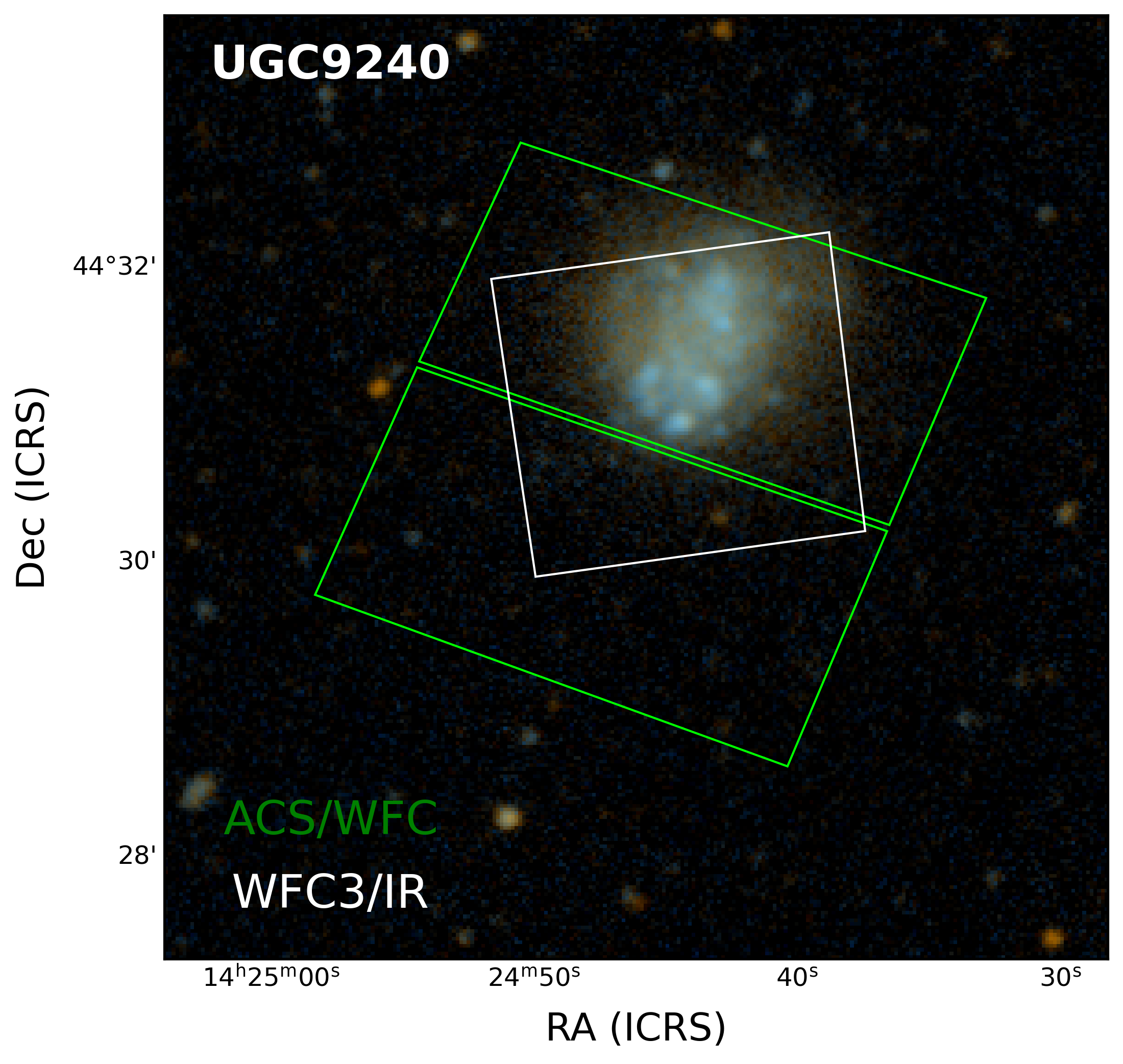}
    \includegraphics[width=0.4\textwidth]{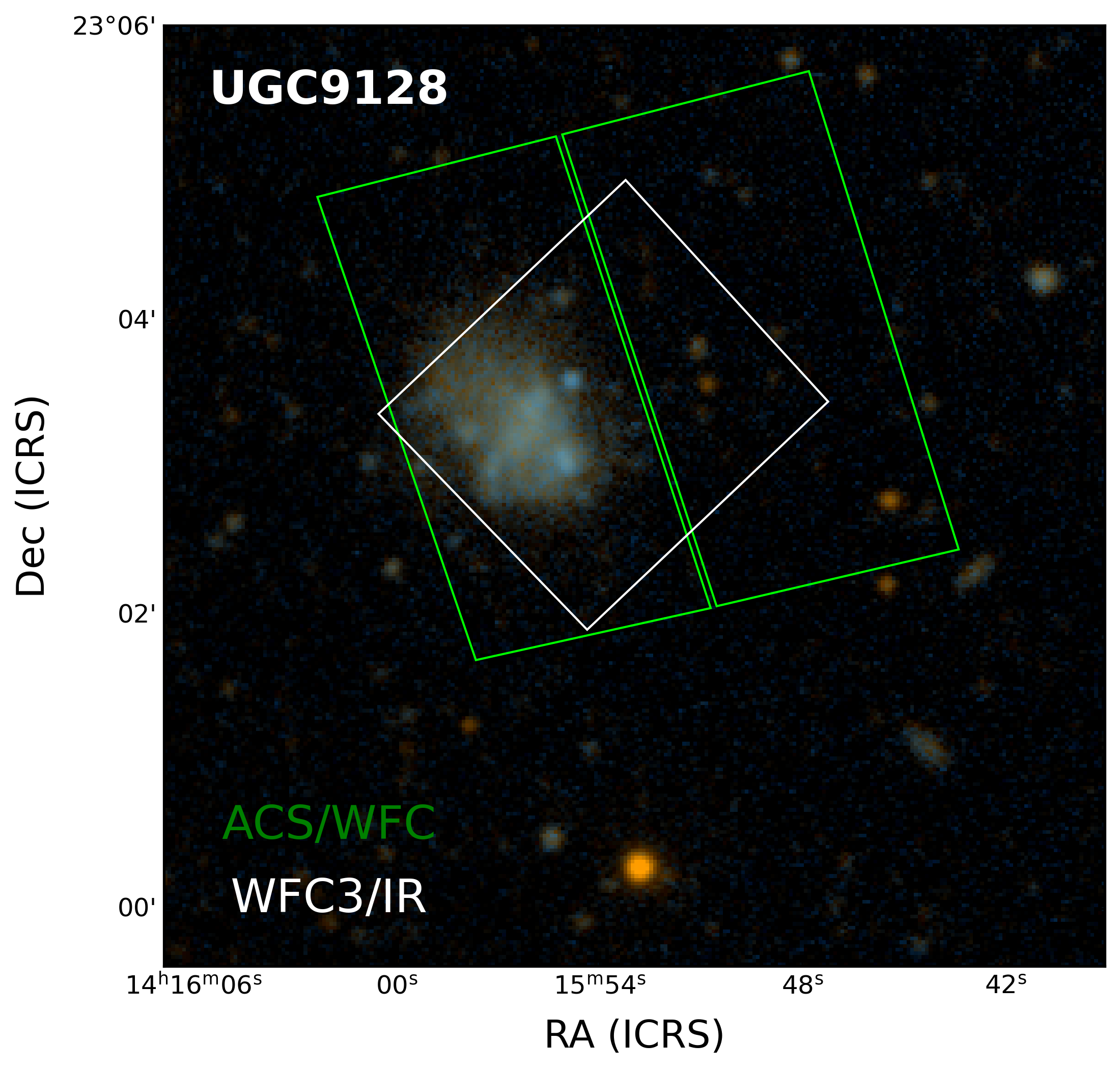}
	\caption{Clockwise from the top left: Antlia Dwarf, UGC8833, UGC9128, and UGC9240. Shown are color images using GALEX near-ultraviolet (NUV; blue) and far-ultraviolet (FUV; red) filters, and an average of the NUV and FUV filters (green), with the HST ACS and WFC3/IR instrument footprints indicated in green and white, respectively. Images are north up, east left. GALEX images were obtained courtesy of NASA/JPL-Caltech.}
	\label{fig:extension_3_color}
\end{figure*}
\end{subfigures}

\subsection{Data Processing:}
\subsubsection{Image Alignment}
Precise, relative astrometric alignment between all images is required for high-fidelity photometry.
We used \TweakReg{} in the  \Drizzlepac{} package \citep{Hack2013, Avila2015} to align our ACS and WFC3/IR raw images to a common world coordinate system (WCS). \TweakReg{} uses a point source detection algorithm to extract astrometric source catalogs for each input image. The alignment algorithm calculates the best fits for shifts, rotations, and scale factors between the images and a reference image based on matching the point sources across images. Images in one filter are first aligned to each other and are then combined into a single drizzled image using \Astrodrizzle{} in \Drizzlepac{}. The drizzled image is then used as a reference image to which images in all other filters are aligned. We selected the drizzled F814W image as the reference image, as this image is a mid-wavelength image in our data set and the ACS data have a higher resolution than WFC3. \TweakReg{} then updates the World Coordinate System (WCS) in the fits header for the non-reference images.

We required the \TweakReg{} alignment to be $\leq 0.1$~pixels for ACS images, and $\leq 0.15$~pixels for WFC3/IR. We used \TweakReg{} to align all the F814W images to each other and then used \Astrodrizzle{} to generate a drizzled F814W reference image. Next, we aligned all remaining CTE-corrected \texttt{flc.fits} ACS images and the \texttt{flt.fits} WFC3/IR images to the drizzled F814W image. Finally, we used \Astrodrizzle{} to make drizzled images in the three other filters (F606W, F110W, and F160W) and visually inspected their alignment using \texttt{DS9} \citep{Joye2003}.

\subsubsection{Multi-Camera Photometry}
To robustly calibrate the TRGB in the NIR we required a catalog of matched stars from all of the data for each galaxy. We performed simultaneous cross-camera point spread function (PSF) photometry on the ACS and WFC3/IR images. PSF photometry on the ACS CTE-corrected and WFC3/IR flat-field calibrated images was executed using \Dolphot{}, a modified version of the WFPC2-specific \texttt{HSTphot} package that includes instrument-specific support for ACS and WFC3 \citep{Dolphin2000, Dolphin2016}. \Dolphot{} uses a deep, cosmic-ray cleaned and drizzled image, produced using \Astrodrizzle{}, as a reference image to identify point sources and alignment solutions. The drizzled F814W was selected as the reference image as ACS has a larger field of view, higher spatial resolution, and a smaller PSF relative to WFC3/IR. \Dolphot{} then locates and iteratively measures the flux of the same point sources in each raw image. 

We adopt the \Dolphot{} input parameters used in the Panchromatic Hubble Andromeda Treasury (PHAT) and Panchromatic Hubble Andromeda: Triangulum Extended Region (PHATTER) surveys \citep{Williams2014,Williams2021}. We conducted extensive input parameter tests by varying several parameters away from those presented in the PHAT survey. We find in particular that for the parameter \texttt{FitSky}, which sets the method \Dolphot{} uses to estimate the local background around each source, a value of 2 produces the best results in both the number and quality of photometered sources in fields with low or high stellar densities. This is in agreement the with the results from the PHATTER survey \citep{Williams2021}.

\subsubsection{Artificial Star Tests}
We performed ASTs using \Dolphot{} in order to quantify photometric completeness and photometric errors due to blending in our stellar catalogs. We injected $200\text{k}-500\text{k}$ ASTs in each field that spanned the CMD parameter space of all four filters (F606W, F814W, F110W, and F160W). We generated the artificial star input lists required to run ASTs using the Bayesian Extinction  And Stellar Tool \citep[\Beast{};][]{Gordon2016}. While \Dolphot{} has the ability to generate artificial star input lists for more than two bands, the artifical star input lists that the \Beast{} generates should be higher-fidelity, since \Dolphot{} uses a simple scaling relation between the bands. Briefly, the \Beast{} uses an observed photometry catalog and a grid of stellar evolution models to predict sets of spectral energy distributions (SEDs). A \Beast{} artificial stars input list is then created by sampling the SEDs ensuring that the full range of the model fluxes in each filter is spanned. 

\subsubsection{Photometric Catalog}
We created high-fidelity stellar catalogs by applying quality cuts to the \Dolphot{} photometric outputs. The initial output from \Dolphot{} includes parameters which provide information about the quality of the recovered points sources. We use a number of these parameters to cull sources that we define as good stars. First, we remove sources with the error flag parameter $\geq 4$. The error flag tells us how well a source was recovered in the image. We also remove sources with the object type parameter $<2$, which tells us about the confidence that a measured object is a star and is not elongated or too faint. Next, we exclude sources with a signal-to-noise ratio (SNR) $\leq5$ in each filter. We then apply cuts based on the crowding parameter, which is a measure of the difference between the initial brightness of a star and its value after subtracting the flux from neighboring sources. We use the sharpness parameter to remove sources that are not point like (i.e., cosmic rays or extended sources such as background galaxies).

\begin{figure*}
	\centering
		\includegraphics[width=\textwidth]{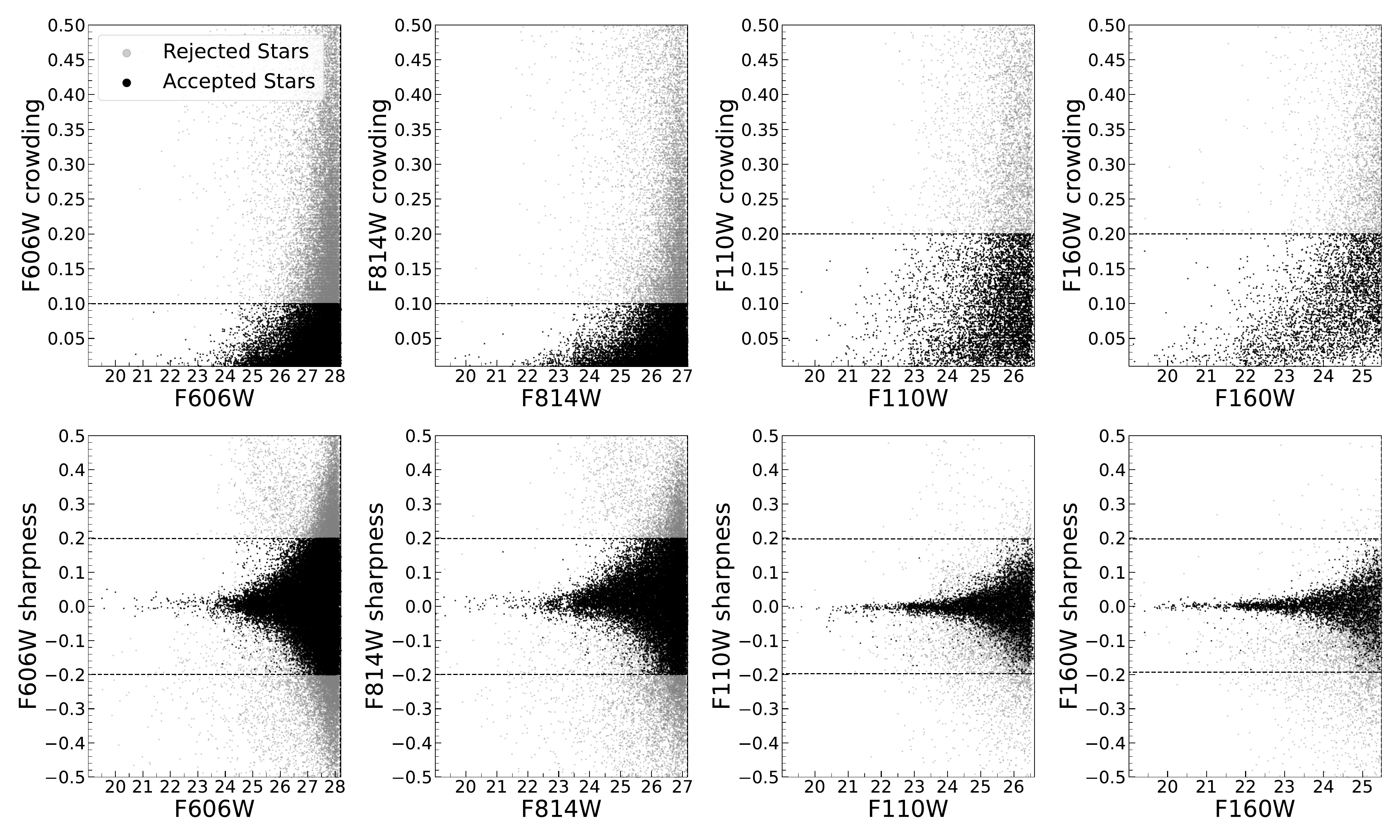}
		\caption{Example of crowding (top) and sharpness (bottom) values as a function of magnitude for NGC2403-2 from the photometry output of \Dolphot{}. From left to right: F606W, F814W, F110W, and F160W filters. Point sources included in our final photometry catalogs are in black and excluded point sources are in gray. Dashed horizontal lines bracket the accepted range of values. 
		}
		\label{fig:quality_cuts}
\end{figure*}

We make our quality cuts empirically and uniformly across all fields, on a per-filter basis. We iteratively adjust the quality parameters until the diffraction spikes from bright foreground stars are minimized in the photometry catalog. Shown in Figure~\ref{fig:quality_cuts} is an example of the criteria used for NGC2403-2,  and is representative of our data. Here we show the stars that pass our quality cuts (black points), and those that fail to meet our criteria (gray points). In F606W and F814W, the maximum crowding value is 0.1 mag, and in F110W and F160W the maximum crowding value allowed is 0.2~mag. In all filters, we chose to include only sources with a value of sharpness in the range $-0.2 \leq\text{sharpness}\leq 0.2$~mag. 

\subsubsection{Color-Magnitude Diagrams}\label{sec:CMDs}
\begin{figure*}[!h]
	\centering
	\includegraphics[width=\textwidth]{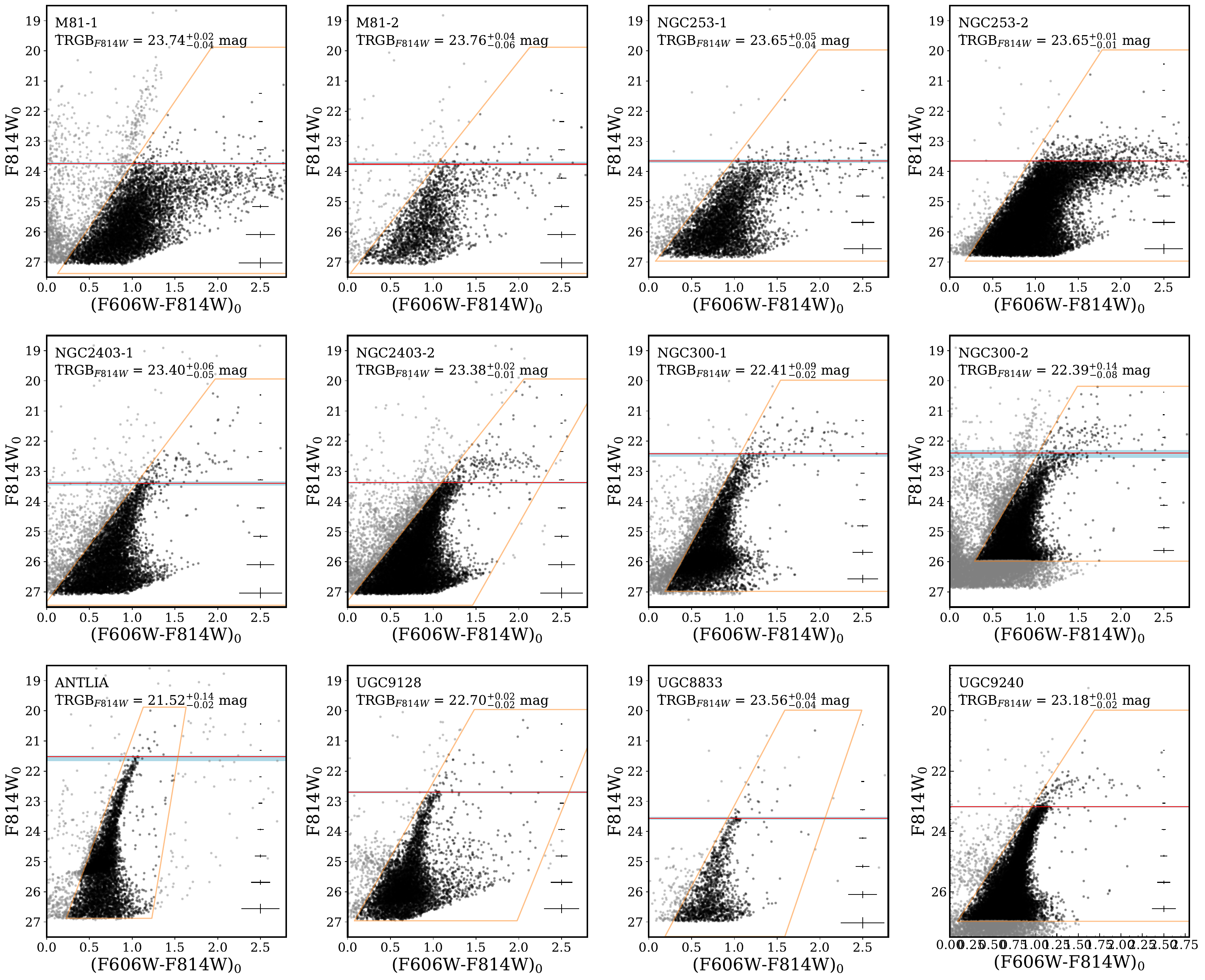}
	\caption{Foreground-extinction corrected optical CMDs with spatial cuts applied to the photometry. From top to bottom in the first and second rows: M81, NGC253, NGC2403, and NGC300. The first and third columns show field 1, while the second and fourth columns show field 2. The third row shows the 4 dwarf galaxies: Antlia Dwarf, UGC9128, UGC8833, and UGC9240. Each panel consists of a CMD using the F606W and F814W filters. Representative photometric uncertainties per magnitude bin are shown. Boxed regions (orange) show the CMD selections to isolate the RGB sequence. We empirically determine the blue edge of box such that it follows the blue edge of the RGB, while the red edge encompasses all sources redward of the blue edge. Point sources inside (outside) the selection box are shown as black (gray) points. The locations of the TRGB magnitudes as identified from the F814W filter with the ML TRGB fit method are shown as  horizontal lines (red) with uncertainties indicated as shaded light blue regions around the fit (see \S~\ref{sec:ACS_TRGB} for more details).}
	\label{fig:original_f814w_trgb}
\end{figure*}

Figure~\ref{fig:original_f814w_trgb} shows extinction corrected optical CMDs from the spatially-clipped ACS FOV with for each galaxy (as shown in green in Figures~\ref{fig:original_3_color} and~\ref{fig:extension_3_color}). Extinction corrections for all the galaxies in the sample were taken from the dust maps of \citet{Schlegel1998} with a re-calibration from \citet{Schlafly2011}. $A_{V}$ values are listed in Table~\ref{tab:table1}. Extinction corrections are small with the maximum values of $A_{\text{F606W}}=0.19$,  $A_{\text{F814W}}=0.12$, $A_{\text{F110W}}=0.069$, and $A_{\text{F160W}}=0.041$ for Antlia Dwarf, and average values for the entire sample of $A_{\text{F606W}}=0.052$, $A_{\text{F814W}}=0.032$, $A_{\text{F110W}}=0.018$, and $A_{\text{F160W}}=0.011$. Table~\ref{tab:reddening} summarizes the filter coefficients and final filter specific total extinction values.

\begin{table*}[]
    \centering
    \begin{tabular}{lcccccc}
    \hline
    \hline
         Target & F606W & F814W & F110W & F160W\\
         & ($R = 2.471$) & ($R = 1.526$) & ($R = 0.881$)& ($R = 0.521$)\\
         & (mag) & (mag) & (mag) & (mag)\\
         \hline
         M81-1 & 0.19 & 0.12 & 0.07 & 0.04\\
         M81-2 & 0.19 & 0.12 & 0.07 & 0.04\\
         NGC253-1 & 0.04 & 0.03 & 0.02 & 0.01\\
         NGC253-2 & 0.05 & 0.03 & 0.02 & 0.01\\
         NGC300-1 & 0.04 & 0.02 & 0.01 & 0.01\\
         NGC300-2 & 0.03 & 0.02 & 0.01 & 0.01\\
         NGC2403-1 & 0.09 & 0.06 & 0.03 & 0.02\\
         NGC2403-2 & 0.10 & 0.06 & 0.03 & 0.02\\
         Antlia Dwarf & 0.19 & 0.12 & 0.07 & 0.04\\
         UGC8833 & 0.03 & 0.02 & 0.01 & 0.01\\
         UGC9128 & 0.06 & 0.03 & 0.02 & 0.01\\
         UGC9240 & 0.03 & 0.02 & 0.01 & 0.01\\
         \hline
    \end{tabular}
    \caption{Foreground total extinction magnitudes for the filters F606W, F814W, F110W, and F160W. The extinction law slope, $R_{\text{filter}}$, is reported for each filter from \citep{Schlafly2011} for an $R_V=3.1$ extinction law.}
    \label{tab:reddening}
\end{table*}

Each galaxy has a well-populated RGB and the TRGB is identifiable by eye as a distinct discontinuity at the upper part of the RGB. We also note the presence of main-sequence, blue helium-burning (BHeB), and red helium-burning (RHeB) stars, as well as some AGB stars in all fields except for NGC253. We select spatial regions in either the stellar halos or outside the dense stellar disks to minimize contamination from these young, non-RGB stellar populations in ACS, as described in Section~\ref{sec:spatial_cuts}.

\begin{figure*}[ht!]
	\centering
	\includegraphics[width=\textwidth]{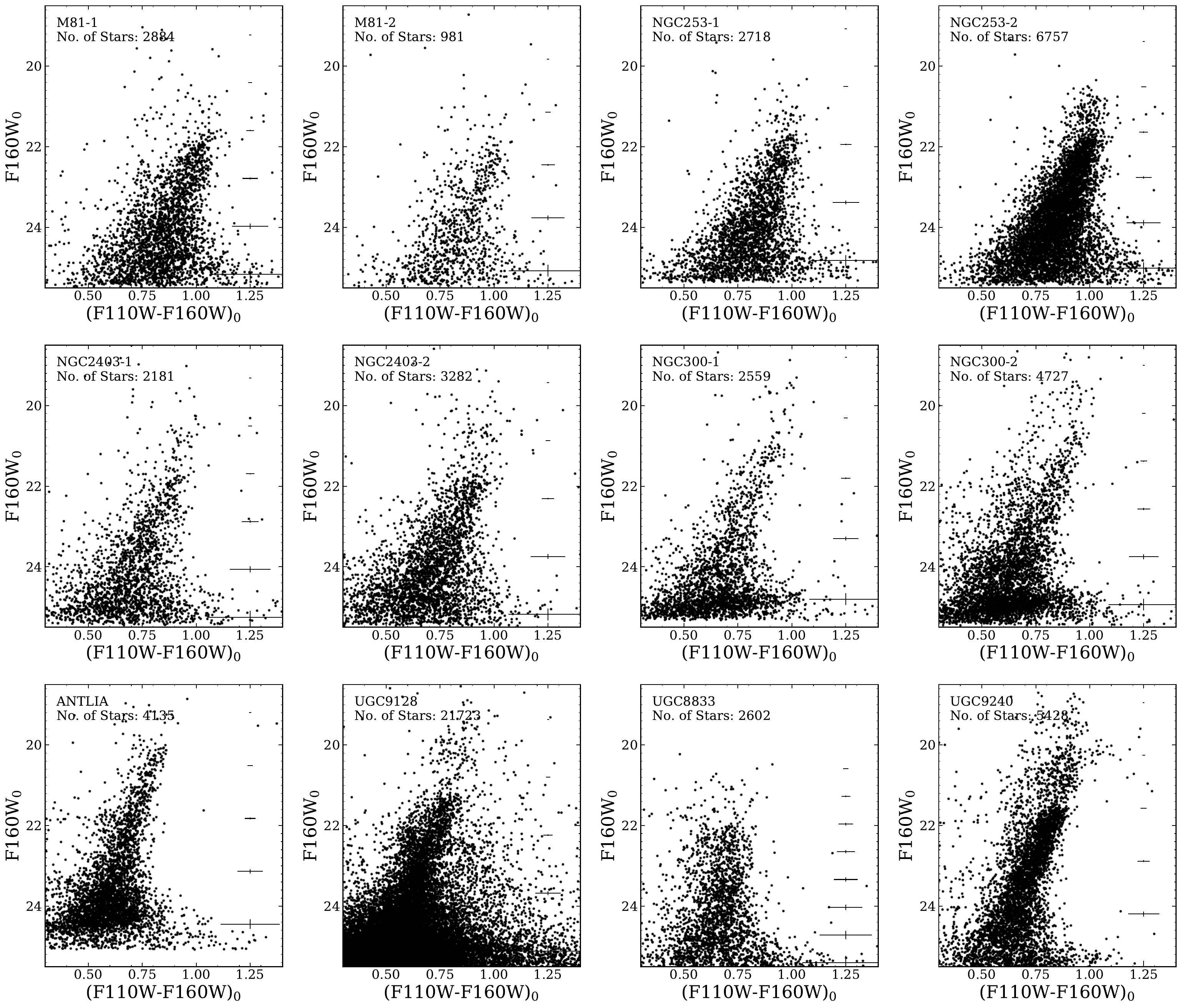}
	\caption{Foreground-extinction corrected NIR CMDs with the F110W and F160W filters. The ordering of CMDs is the same as in Figure~\ref{fig:original_f814w_trgb}. The star counts in each overlapping ACS-WFC3/IR region are provided in the legends. Representative photometric uncertainties per magnitude bin are shown.}
	\label{fig:wfc3_cmds}
\end{figure*}

Figure~\ref{fig:wfc3_cmds} presents extinction corrected NIR CMDs from the full WFC3/IR FOV for each galaxy (as shown in white in Figures~\ref{fig:original_3_color} and~\ref{fig:extension_3_color}). All NIR CMDs display a similar RGB sequence but the RGB has a narrower color baseline and the TRGB is more steeply sloped compared to the optical CMDs as predicted from stellar isochrones \citep[e.g., ][]{McQuinn2019}. 
\\
\subsubsection{Spatial Cuts}\label{sec:spatial_cuts}
The TRGB is best identified in a CMD with a well-populated RGB sequence and minimal contamination from non-RGB stars (i.e., AGB or RHeB stars). Because the TRGB is based on measuring the discontinuity in the LF, retention of non-RGB stars in the photometry with both colors and apparent brightnesses near the RGB can blur the break in the LF.

Thus, we apply spatial cuts to our observations to reduce contamination from non-RGB populations. Our spatial cuts are based on a 2-step approach and use information from the ACS stellar catalogs and the location of the ACS FOV relative to the main stellar disks of the galaxies. First, we exclude regions where we find evidence of significant young stellar populations based on GALEX images (see Figures~\ref{fig:original_3_color} and \ref{fig:extension_3_color}). Second, we refine the spatial cuts by making CMD selections of any remaining young stellar populations, plotting their spatial distributions, and excluding any region with high fractions of young stars. 

This 2-step approach is modified for the four dwarf galaxies with archival observations in which the ACS FOV covers the central region and outer stellar field. We adopt ellipse structural parameters from \citet{McConnachie2012} for Antlia Dwarf, Sloan Digital Sky Survey Data Release 16 \citep[SDSS DR16; ][]{Ahumada+2020} for UGC~8833, and SDSS DR12 \citep{Alam2015} for UGC~9128 and for UGC~9240. We then generate elliptical annuli starting from the center and moving outward, and count the number of MS stars and RGB stars in each annuli. We use the ratio of MS stars to RGB stars to optimize our final FOV selection.

\begin{figure*}
	\centering
		\includegraphics[width=\textwidth]{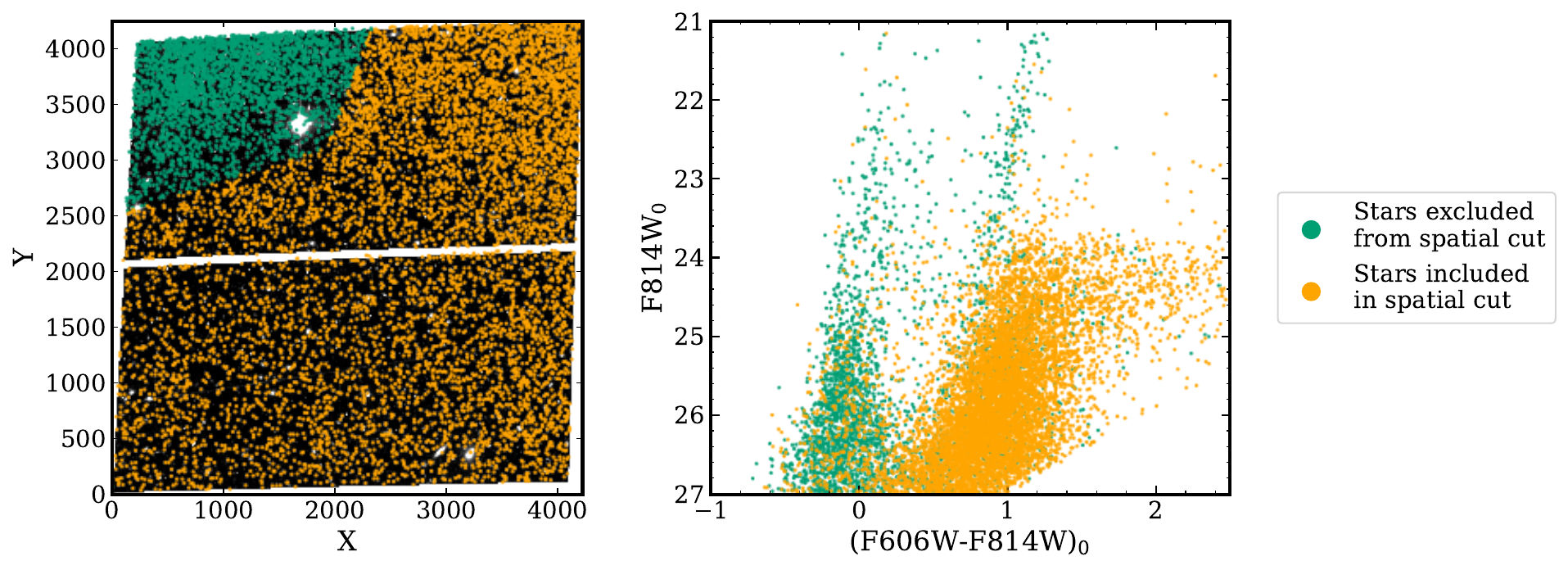}
		\caption{Example of spatial cuts that were carefully applied to specific fields. Left: M81-1 ACS color image with point sources from our final photometry catalog overlaid. Stars in the upper left corner of the image (green points) are excluded from the final photometry catalog used to fit the TRGB. The remaining stars (orange points) are kept after our spatial cuts. Right: Extinction corrected optical CMD of the stars from the spatially excluded region (green) and spatially included region (orange). The spatial cut removes most main-sequence stars and a significant fraction of the HeB sequences, while retaining the majority of the RGB stars.}
		\label{fig:spatial_cut_demo}
\end{figure*}

The left panel of Figure~\ref{fig:spatial_cut_demo} presents an example of the spatial cuts applied to M81-1. The point sources in the upper left corner of the image (green) are separated from the rest of the sources (orange). The right panel presents the resulting CMD of the two regions color-coded in the same manner. There is a clear separation of the stellar populations in the these two regions. 

\section{Measuring the Distances in the F814W Luminosity Function}\label{sec:ACS_TRGB}
In this section, first we introduce the calibration of the F814W TRGB magnitude as a function of the F606W$-$F814W color. Second, we discuss two of the methods developed to measure the magnitude of the TRGB, and justify the method we employ to measure the TRGB in our sample. Third, we adopt our F814W TRGB zero point and we present our final F814W-based distance measurements. Finally, we compare our results to those reported in the literature.

\subsection{Color Correction of the TRGB in F814W}\label{sec:trgb_opt_measure}
The first step taken to identify the TRGB luminosity is to correct for the well-known, modest dependence of the TRGB on metallicity/age. We account for this dependence by using color-based corrections. We adopt the only the color correction (not the zero point) from \citet{Jang2017}, named the quadratic transformation (QT) system, derived for the HST filters which has been recently shown as most accurately characterizing the color-based dependence of the TRGB \citep{Hoyt2021}. We reproduce the calibration here, over the applicable color range F606W-F814W$>1.5$, for convenience:

\resizebox{1.0\linewidth}{!}{
	\begin{minipage}{1.05\linewidth}
        \begin{eqnarray}\label{eqn:Jang2017}
            \text{F814W}_{\text{rect}} = \text{F814W} 
            &-\alpha\left[\left(\text{F606W-F814W}\right)-1.1\right]^2 \\\nonumber
            &-\beta\left[\left(\text{F606W-F814W}\right)-1.1\right] \\ \nonumber
        \end{eqnarray}
\end{minipage}
}
\\ \\ 
where F814W is the original magnitude, F814W$_{\text{rect}}$ is the color-corrected magnitude, and $\alpha = 0.159\pm0.010$ and $\beta=-0.047\pm0.020$ are the best-fit parameters for the quadratic and linear terms, respectively. For F606W-F814W$\leq 1.5$, the TRGB is consistent with being flat (i.e., at a constant magnitude) and no correction is applied.
In Eq.~3 of \citet[][ Eq.~\ref{eqn:Jang2017} in this text]{Jang2017} the authors identify the pivot point for the color-based correction as the average color of the TRGB (F606W$-$F814W $=1.1$~mag) in their sample. The use of the average color as the pivot point accounts for differences between the color of the stars in the observed galaxy and the color used to anchor the zero point in the calibration. We apply this correction on a per-star basis prior to measuring the location of TRGB (i.e., we rectify the TRGB). Fitting for the TRGB after applying the correction has the effect of increasing the sharpness of the discontinuity in the LF. Note that this color-correction only applies to 2 of the 12 fields as most of the data had colors F606W-F814W$<1.5$. 

\subsection{TRGB Measurements Methods}
Several techniques have been developed to measure the F814W TRGB magnitude. Of them, the technique first applied to identify the discontinuity in the LF was the Sobel filter edge detection technique \citep{Lee1993, Sakai1996, Sakai1997}. This technique is based on applying a Sobel Kernel in the form $[-1, 0, +1]$, an approximation of a derivative, to a binned LF. Convolving the Sobel Kernel with the binned LF results in a response function indicating the greatest change in number of stars from one magnitude bin to the next.

We employ a more sophisticated Bayesian Maximum Likelihood (ML) technique to measure the F814W TRGB magnitude in each galaxy. The ML technique uses a parametric form of the RGB LF to fit the observed LF.  The probability distribution takes into account the photometric uncertainty distribution and completeness function from ASTs \citep[see][for a full discussion of the technique, and for a detailed discussion of quantifying tip uncertainties respectively]{Makarov2006,Madore2023}. We adopt the same theoretical LF form used in \citet{Makarov2006}:

\begin{equation}
P = 
\begin{cases}
	10^{\left(A*\left(m-m_{\text{TRGB}}\right)+B\right)},  & \text{if } m-m_{\text{TRGB}} \geq 0 \\
	10^{\left(C*(m-m_{\text{TRGB}})\right)},  & \text{if } m-m_{\text{TRGB}} < 0 
\end{cases}
\label{eq:ml_form}
\end{equation}
where A is the slope of the RGB with a normal prior of $\mu_{\text{A}}=0.3$  and uncertainty $\sigma_{\text{A}}=0.07$, B is the RGB jump, and C is the AGB slope with a normal prior of $\mu_{\text{C}}=0.3$ and uncertainty $\sigma_{\text{C}}=0.2$. The parameters A, B, and C are all free parameters. The ML measurement uncertainty is based on the range of solutions returning the log of the probability, $\log{(P)}$, within 0.5 of the maximum. 

When fitting for the TRGB in the F814W filter, we select stars consistent with the luminosity and colors of RGB stars on a per-field basis as a well constrained RGB sequence will yield the most robust TRGB measurements. If stars blueward of the RGB sequence (e.g., BHeB, RHeB, or main sequence stars) remain in the CMD, they can bias the stellar LF and reduce the strength of the break in the LF associated with the TRGB. The color ranges we selected to constrain the RGB stars are shown in Figure~\ref{fig:original_f814w_trgb} as orange boxes with stars included as black points, while the stars excluded are shown as gray points. Our color selections are empirically determined such that the blue edge of the box follows the blue edge of the RGB, while the red edge is set to encompass the reddest sources. Even though the ML approach takes into account the possible presence of a secondary population, we nonetheless explored restricting our fits to stars within a red color limit (thereby reducing the potential number of redder AGB stars). When we compared the TRGB values measured without applying this additional color restriction, we find the residuals between the fits are consistent with zero within the statistical uncertainties. We chose to use the full population of redder stars as this selection method avoids artificially biasing the AST-based completeness function used in the ML fitting.

We run the ML technique on the foreground-extinction corrected photometry, after applying the color-based correction described in Eq.~\ref{eqn:Jang2017}. The location of the TRGB magnitude identified with the ML technique (indicated as a horizontal red line in Figure \ref{fig:original_f814w_trgb}) is shown in each CMD with uncertainties from the ML fit shaded in blue. Table \ref{tab:table3} presents the best-fit TRGB luminosities from the ML technique for the sample. The uncertainties include statistical uncertainties from the ML probability distribution function.

This ML technique is preferred over the Sobel filter technique for measuring the F814W TRGB magnitude as it does not require binning the LF, and considers the photometric measurement uncertainties and completeness. The ML technique also fits the F814W LF while accounting for the AGB population. Compared with the Sobel filter technique, the ML uncertainties are typically lower, primarily due to the Sobel filters dependence on bin width. In Section~\ref{sec:NIR_TRGB_Calibration}, however, we employ the Sobel filter technique to measure the slope and zero point of the NIR TRGB for two reasons. First, we perform our calibration based on a stacked stellar catalog of every field in the sample. As of this writing, the ML method does not have a way to simultaneously handle the results of ASTs from multiple fields. It assumes any input photometry is from one field, and that the ASTs characterize the photometric noise properties of that field only.
Second, the Sobel filter proves to be an excellent tool to optimize and ultimately determine our best-fit slope to the TRGB. 

\subsection{Distances}\label{sec:acs_distances}
We use the zero point calibration from \citet{Freedman2021} to convert TRGB luminosities to distance moduli  ($\mu$). Their final F814W zero point is based on four independent calibrations of the TRGB absolute magnitude: the megamaser galaxy NGC~4258, Galactic Globular clusters (i.e., using Gaia DR3), the Large Magellanic Cloud (LMC), and Small Magellanic Cloud (SMC). The independent calibrations are found to be internally self-consistent at the $1\%$ level. We therefore choose to adopt this F814W TRGB zero point measured. We note that, since our NIR TRGB calibration is tied to the F814W TRGB calibration, the results of our calibration can be updated to account for improvements to the F814W TRGB zero point and uncertainties.
 
 Statistical uncertainties from the ML fit are reported as the total uncertainty on the TRGB measurement (see \S\ref{sec:error_budget} for details). Final distance moduli and distances are also reported in Table \ref{tab:table3} with statistical uncertainties on each fit. For the four galaxies with two observed fields, the best-fitting distances agree within the uncertainties.

\begin{table*}[t!]
	\centering
	\footnotesize
	\caption{Summary of Distances Measured from F814W LF}
	\label{tab:table3}
	\begin{tabular}{l | ccc}
	\hline
	Galaxy  & m$_{\text{F814W},0}$                         & Distance Modulus           & Distance \\ 
			&                                   & (mag)		                 & (Mpc) \\
	\hline
	\hline
	M81-1 &	23.74$^{+0.02}_{-0.04}$ & 27.79$^{+0.02}_{-0.04}$ & 3.61$^{+0.04}_{-0.07}$ \\
M81-2 &	23.76$^{+0.04}_{-0.06}$ & 27.80$^{+0.04}_{-0.06}$ & 3.64$^{+0.06}_{-0.10}$ \\
NGC253-1 &	23.65$^{+0.05}_{-0.04}$ & 27.70$^{+0.05}_{-0.04}$ & 3.46$^{+0.08}_{-0.06}$ \\
NGC253-2 &	23.65$^{+0.01}_{-0.01}$ & 27.70$^{+0.01}_{-0.01}$ & 3.46$^{+0.02}_{-0.02}$ \\
NGC300-1 &	22.41$^{+0.09}_{-0.02}$ & 26.46$^{+0.09}_{-0.02}$ & 1.96$^{+0.08}_{-0.02}$ \\
NGC300-2 &	22.39$^{+0.14}_{-0.08}$ & 26.44$^{+0.14}_{-0.08}$ & 1.94$^{+0.13}_{-0.08}$ \\
NGC2403-1 &	23.40$^{+0.06}_{-0.05}$ & 27.45$^{+0.06}_{-0.05}$ & 3.09$^{+0.09}_{-0.08}$ \\
NGC2403-2 &	23.38$^{+0.02}_{-0.01}$ & 27.43$^{+0.02}_{-0.01}$ & 3.06$^{+0.03}_{-0.02}$\\
Antlia Dwarf & 21.52$^{+0.14}_{-0.02}$ & 25.57$^{+0.14}_{-0.02}$ & 1.30$^{+0.08}_{-0.01}$\\
UGC8833 &	23.56$^{+0.04}_{-0.04}$ & 27.61$^{+0.04}_{-0.04}$ & 3.33$^{+0.05}_{-0.06}$\\
UGC9128 &	22.70$^{+0.02}_{-0.02}$ & 26.75$^{+0.02}_{-0.02}$ & 2.24$^{+0.02}_{-0.02}$ \\
UGC9240 &	23.18$^{+0.01}_{-0.02}$ & 27.23$^{+0.01}_{-0.02}$ & 2.80$^{+0.02}_{-0.03}$\\
    
	\hline \hline
	\multicolumn{4}{l}{
		\begin{minipage}{8cm}~\\
			NOTE - The apparent magnitudes reported $\left(m_{\text{F814W,0}}\right)$ in column 1 are measured after applying the color correction from \citet{Jang2017}. We adopt the ACS F814W TRGB zero point from \citet{Freedman2021} for our distance moduli, $M^{\text{TRGB}}_{\text{F814W}} =-4.049\pm0.015$~mag. The reported uncertainties include uncertainties from the F814W TRGB ML fits.
		\end{minipage}
	}
	\end{tabular}
\end{table*}

\subsection{Comparison with TRGB Distances from the Literature}\label{sec:Distance_Comparison}
Since we require robust F814W TRGB-based distances to anchor the NIR TRGB calibration, we compared distance moduli between our measurements and TRGB-based measurements in the literature. Based on the values reported in the NASA/IPAC Extragalactic Database (NED)\footnote{The NASA/IPAC Extragalactic Database (NED) is funded by the National Aeronautics and Space Administration and operated by the California Institute of Technology.} our measurements agree within 1$\sigma$ of the mean for each target. Note that these methods include slightly different approaches to measuring and calibrating the TRGB.

We included a more direct test and focus on TRGB measurements from the literature that match our approach. Specifically, we selected distance moduli values from the literature sources to compare with our data based on the following criteria: TRGB-based distance measurements were made using HST F814W data and color-based corrections, and zero points used in the measurement were empirical. We compared the individual measurements to our measurements. Again, we found that the measurements agree within their uncertainties. These literature values are tabulated in Table~\ref{tab:table4}.

\begin{table*}[t!]
	\centering
	\footnotesize
	\caption{F814W TRGB-based Comparison of Distance Moduli}
	\label{tab:table4}
	\begin{tabular}{l|ccl}
		\hline
		Galaxy & $\mu_{\text{F814W}}$ (This Study) & 	$\mu_{\text{F814W}}$ (Literature) & References \\
        &   (mag) & (mag) & \\
		\hline \hline
		M81-1 &	27.79$^{+0.02}_{-0.04}$ & $27.79\pm0.07$& \citet{Radburn-Smith2011} \\
		M81-2 &	27.80$^{+0.04}_{-0.06}$ & -- & -- \\
		NGC253-1 &	27.70$^{+0.05}_{-0.04}$ & $27.70\pm0.07$ & \citet{Radburn-Smith2011}\\
		NGC253-2 &	27.70$^{+0.01}_{-0.01}$ & -- & -- \\
		NGC300-1 &	26.46$^{+0.09}_{-0.02}$ & $26.60^{+0.06}_{-0.05}$& \citet{Jacobs2009}\\
		NGC300-2 &	26.44$^{+0.14}_{-0.08}$ & -- & --  \\
		NGC2403-1 &	27.45$^{+0.06}_{-0.05}$ & $27.51\pm0.07$ & \citet{Radburn-Smith2011} \\
		NGC2403-2 &	27.43$^{+0.02}_{-0.01}$ & -- & -- \\
		Antlia Dwarf &	25.57$^{+0.14}_{-0.02}$ & $25.68^{+0.09}_{-0.10}$ & \citet{Jacobs2009}\\
		UGC8833 & 27.61$^{+0.04}_{-0.04}$ & $27.56^{+0.06}_{-0.05}$ & \citet{Jacobs2009} \\
		UGC9128 & 26.75$^{+0.02}_{-0.02}$ & $26.81\pm0.04$ & \citet{Jacobs2009}\\
		UGC9240 &	27.23$^{+0.01}_{-0.02}$ & $27.26^{+0.04}_{-0.03}$ & \citet{Jacobs2009}\\
		\hline\hline
		\multicolumn{4}{l}{
		\begin{minipage}{12cm}~\\
			NOTE - We report only TRGB-based distance moduli values from the literature that were made using HST F814W data, and that used empirical color-based corrections and zero points in the measurement. Both \citep{Jacobs2009} and \citep{Radburn-Smith2011} adopt the TRGB calibration from \citep[][; the most current calibration at the time of their studies]{Rizzi2007}. While their adopted calibration differs from ours, the methodology remains the similar.
		\end{minipage}
	}
	\end{tabular}
\end{table*}

\section{Identifying and Calibrating the TRGB in the NIR}\label{sec:NIR_TRGB_Calibration}
While the TRGB magnitude in the F814W filter is observed to have a modest dependence on metallicity, the NIR TRGB is known to have a stronger dependence. Specifically, it has been shown that the brightness of the TRGB increases with increasing metallicity in the NIR \citep{Valenti2004, Salaris2005, McQuinn2019}. Observationally, this effect manifests as the TRGB luminosity increasing at redder colors. 

To calibrate the slope and zero point of the NIR TRGB in the F110W and F160W filters we made a combined stellar catalog from the entire sample to provide a broad metallicity, hence color, baseline. We found that the stars in a CMD of any individual field did not cover a wide enough color range to robustly characterize the slope of the NIR TRGB for calibration purposes. Additionally, by combining the individual catalogs we increased the total number of stars in the CMD. Shown in the top panels of Figure~\ref{fig:slope_zp_best_corner_w_uncertainties} are the IR CMDs generated from the combined stellar catalogs. The combined stellar catalog is created after converting all foreground-extinction corrected apparent magnitudes to absolute magnitudes using the distance moduli measured from the F814W TRGB listed in Table~\ref{tab:table3}. 

Here, we provide a mathematical basis for the TRGB calibration, introduce our method for identifying the TRGB in the F110W and F160W filters, and then describe our methods for measuring and calibrating its slope as a function of color, and then for measuring the zero point of the NIR TRGB in both filters.

\subsection{Mathematical Background on the TRGB Characterization}\label{sec:calib_math_basis}
The magnitudes of TRGB stars in the NIR exhibit an approximately linear relationship with color such that their brightness increases from bluer to redder colors. Like the F814W TRGB, the color dependence of the NIR TRGB can be calibrated via a color-based correction \citep[e.g.,][]{Rizzi2007, Wu2014, Jang2017}. By characterizing its color dependence, the NIR TRGB can be rectified to a single magnitude and subsequently employed as a robust distance indicator. 

We characterize the color-based correction via the parameter $\beta$ (i.e., the slope of the NIR TRGB), the median F110W-F160W color of the TRGB stars, which sets the pivot point at which the TRGB is rectified to a single magnitude (i.e., the zero point), and the absolute magnitude of the TRGB at the pivot point. We adopt for our color-based correction the following equation:

\begin{align}\label{eqn:color_correction}
	M^{\text{TRGB}}_{\text{NIR}} &= M^{\text{TRGB}}_{\text{NIR},\gamma} - \\
 &\quad \beta\left[\left(\text{F110W-F160W}\right)-\gamma\right] \nonumber
\end{align}
where $\beta$ is the slope of the TRGB in the chosen NIR filter, F110W-F160W is the color of the individual stars, and $\gamma$ is the median F110W-F160W color of the TRGB stars, $M_{\text{NIR}}$ is the color-corrected TRGB magnitude, and $M^{\text{TRGB}}_{\text{NIR},\gamma}$ is the magnitude of the TRGB at the pivot point in the NIR filters. When applied to the apparent magnitude of stars the equation is:
\begin{equation}
\small
    m_{\text{NIR,rect}} = m_{\text{NIR,orig}} - \beta\left[\left(\text{F110W-F160W}\right)-\gamma\right]\\
\end{equation}
where $m_{\text{NIR,rect}}$ and $m_{\text{NIR,orig}}$ are the extinction corrected rectified and original magnitudes of the stars in a NIR filter.

Mathematically, one can transform directly from the calibration in one bandpass, F110W for example, directly to another, F160W, for the same color F110W-F160W \citep[e.g.,][]{Wu2014,Serenelli2017,Madore2020}. The transformation can also be applied between different bandpasses \citep[see ][]{Madore2020}.
Briefly, for the filters F110W and F160W and using Eq.~\ref{eqn:color_correction}:
\begin{align}\label{eqn:F110W_linear_correction}
\text{F110W} =& M^{\text{TRGB}}_{\text{F110W},\gamma} \\
& \quad + \beta\left[\left(\text{F110W-F160W}\right)-\gamma\right]\nonumber
\end{align}
where $M^{\text{TRGB}}_{\text{F110W},\gamma}$ is the absolute magnitude in F110W at the pivot point $\gamma$, and $\beta$ is the slope of the TRGB in F110W. We can then determine the transformation from F110W to F160W. 
First we define F160W as:
\begin{align}
    \text{F160W} = \text{F110W} -(\text{F110W-F160W})
\end{align}
Next we substitute in for F110W with Eq.~\ref{eqn:F110W_linear_correction}.
\begin{align}
    \text{F160W} &= M^{\text{TRGB}}_{\text{F110W},\gamma}\\
    &\quad+ \beta\left[\left(\text{F110W-F160W}\right)-\gamma\right]\nonumber\\ 
    &\quad- (\text{F110W-F160W})\nonumber
\end{align}
then expand the terms on the right hand side,
\begin{align}
&=M^{\text{TRGB}}_{\text{F110W},\gamma}\\
&\quad+ \beta\left(\text{F110W-F160W}\right) -\beta\gamma\nonumber\\
&\quad- (\text{F110W-F160W})\nonumber
\end{align}
combine like terms and expand $\beta\gamma$,
\begin{align}
    &=M^{\text{TRGB}}_{\text{F110W},\gamma}\\
    &\quad+ \left(\text{F110W-F160W}\right)\left(\beta-1\right)\nonumber\\
    &\quad -\left(\beta-1\right)\gamma -\gamma\nonumber
\end{align}
and combine like terms once more to reach the calibration equation for F160W in the form of Eq.~\ref{eqn:F110W_linear_correction}.
\begin{align}
    \text{F160W}&= \left(M^{\text{TRGB}}_{\text{F110W},\gamma}-\gamma\right)\\
    &\quad+ \left(\beta-1\right)\left[\left(\text{F110W-F160W}\right)-\gamma\right]\nonumber
\end{align}
The resulting transformation from F110W to F160W can then be characterized by the coefficients 
\begin{align}
    &M^{\text{TRGB}}_{\text{F160W},\gamma} = M^{\text{TRGB}}_{\text{F110W},\gamma}-\gamma \label{eqn:transform_eq_zp} \\
    &\beta_{\text{F160W}}=\beta_{\text{F110W}}-1. \label{eqn:transform_eq_slope}
\end{align}
While it a straightforward process to transform the parameters $\alpha$, $\beta$, and $M^{\text{TRGB}}_{\text{NIR}}$ from F110W to the F160W, we chose to instead calibrate each filter independently and only then use the transformation as a check on our empirical results. The final values we report for $M^{\text{TRGB}}_{\text{F110W},\gamma}$, $M^{\text{TRGB}}_{\text{F160W},\gamma}$, and $\beta$ are derived independently and empirically from individual fits to the data (see \S\ref{sec:slope_and_zp_measure}). Running the calibrations separately for each filter enables us to consider the photometric uncertainties and their impact on the calibration independently. We then apply the mathematical transformation to the F110W best-fit calibration values as a consistency check to determine the theoretical calibration parameter values in F160W (see \S\ref{sec:error_budget}).

\subsection{NIR TRGB Color Dependence and Zero Point}\label{sec:slope_and_zp_measure}
Our method for calibrating the color-based correction for the NIR TRGB simultaneously measures the slope and zero point of the TRGB. First, we identify a fiducial slope by eye in the NIR CMDs and use that value as a prior in the analysis. We make CMD cuts in the range of the RGB to reduce non-RGB contamination blueward of the RGB, similar to our method for constraining the RGB in the F814W filter. We identify the median color of the TRGB within a color range $0.73\leq\text{F110W-F160W}\leq1.0$~mag at $\gamma=0.92$~mag.

Second, we employ an algorithm based on the Nelder-Mead formalism via \texttt{Scipy's} minimization routine. We construct a binned LF from the rectified CMD within the defined TRGB color range using a bin width of $0.01$~mag and smooth over the LF with the Gaussian-weighted locally estimated scatterplot smoothing (GLOESS) method \citep{Persson2004, Monson2017}. GLOESS is a non-parametric weighted least squares method in which the fitting is done locally and the weights are described by a Gaussian profile with the bandwidth of the kernel described by $\tau$. We choose a smoothing scale of $\tau=0.05$ (see \S~\ref{sec:error_budget} for the systematic uncertainty associate with the choice of smoothing scale). We then apply the GLOESS-smoothed LF with a Sobel filter kernel of width [-1,0,1] convolved with a weighting that suppresses Poisson noise \citep[e.g.,][]{Madore2009, Hatt2017}. The minimization routine explores the slope parameter space to maximize the Sobel response at the TRGB magnitude under the assumption that the largest Sobel response corresponds to the slope best characterizing the TRGB color dependence. We bound the slope values to ranges of $(-1.7, -0.6$) and $(-2.8, -1.5)$ for F110W and F160W, respectively, use a tolerance of $1\times10^{-6}$ for the Nedler-Mead algorithm, and search for the Sobel peak response and corresponding TRGB magnitude over a range of $\pm0.2$~mag about the expected location of the TRGB.

Third, we quantify the best-fit slope, zero point, and their uncertainties by running 1500 Monte Carlo (MC) simulations. Each MC simulation repeats step two for resampled magnitudes of the stars based on the uncertainties in the $\mu_{\text{F814W}}$ measurements and on the photometric uncertainties determined from ASTs. The latter uncertainties are determined by finding the 2000 recovered artificial stars closest (e.g., $d=\sqrt{\text{magnitude}^2 + \text{color}^2}$) to each observed star and calculating the standard deviation and median (i.e., the bias) of the difference in their recovered and input magnitudes. The bias $\pm0.1$~mag about the TRGB magnitudes in F110W and F160W is $<0.001$ mag. In addition, we resample the colors of the stars based on the photometric uncertainties in the two NIR filters. We measure the final slope and zero point values at the $50$th percentile of the MC output, and measure asymmetric uncertainties by taking the differences in the $16$th and $84$th percentiles from the median. We provide the following relationships in the WFC3/IR Vegamag system for the absolute magnitude calibrations in the F110W and F160W filters:
\begin{align}\label{eqn:f110w_color_correction}
\small
    M^{\text{WFC3}}_{\text{F110W}} = \bluezp\blueslope\left[\left(\text{F110W-F160W}\right)-0.92\right]
\end{align}
\begin{align}\label{eqn:f160w_color_correction}
\small
    M^{\text{WFC3}}_{\text{F160W}} = \redzp\redslope\left[\left(\text{F110W-F160W}\right)-0.92\right]
\end{align}

\begin{figure*}[!tbh]
	\centering
	\includegraphics[width=0.35\textwidth]{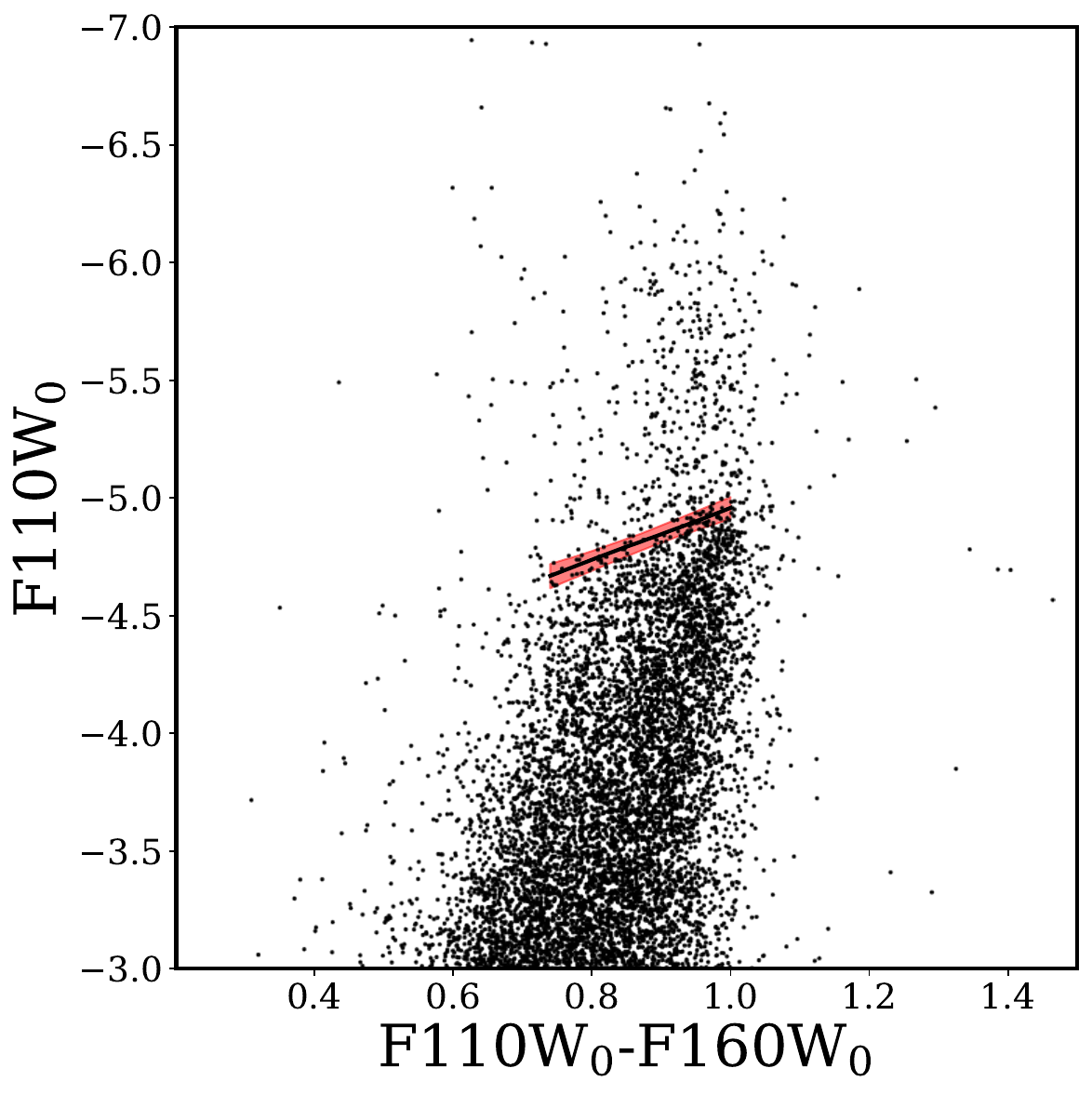}
    \includegraphics[width=0.35\textwidth]{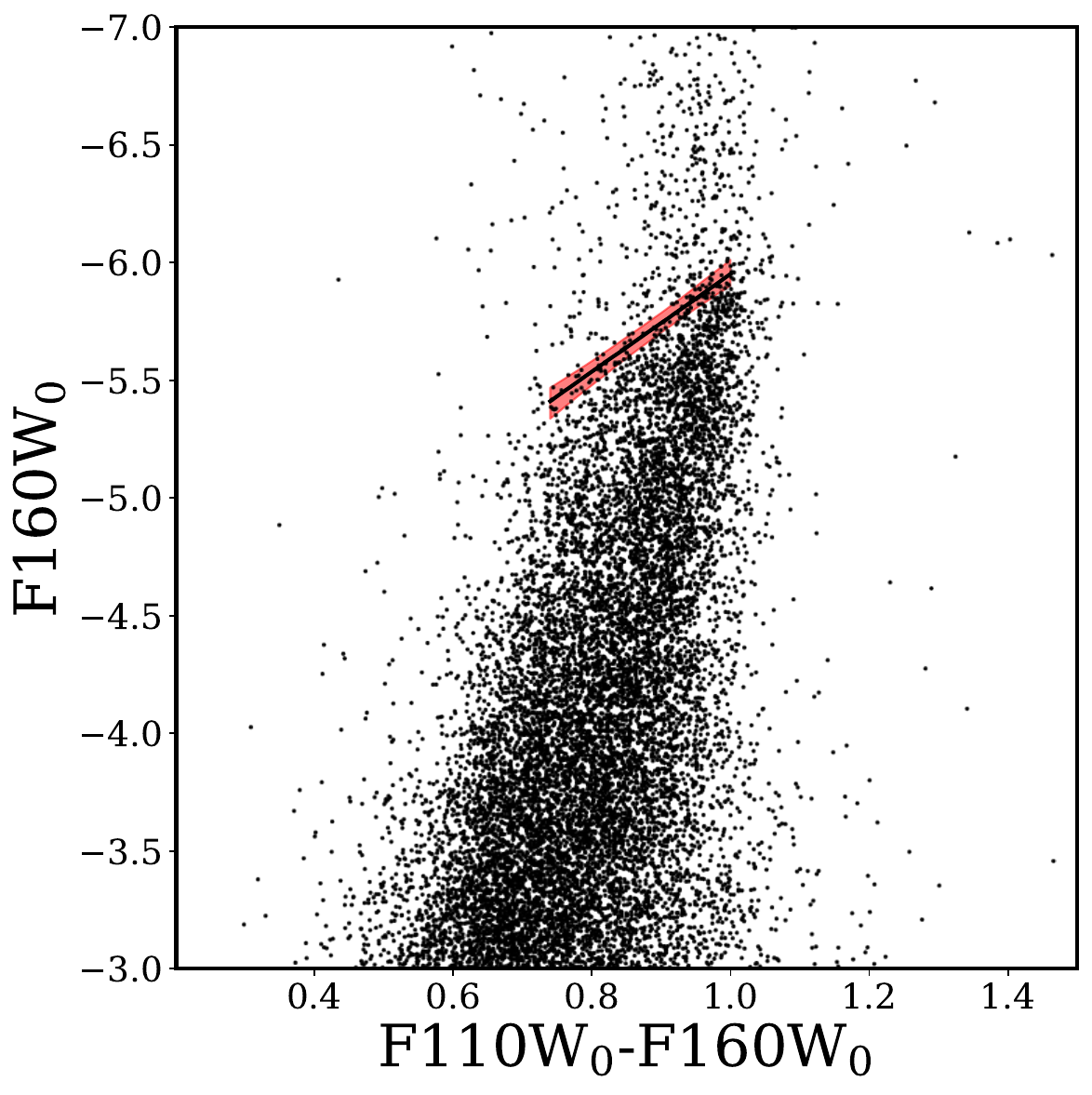}
    \includegraphics[width=0.35\textwidth]{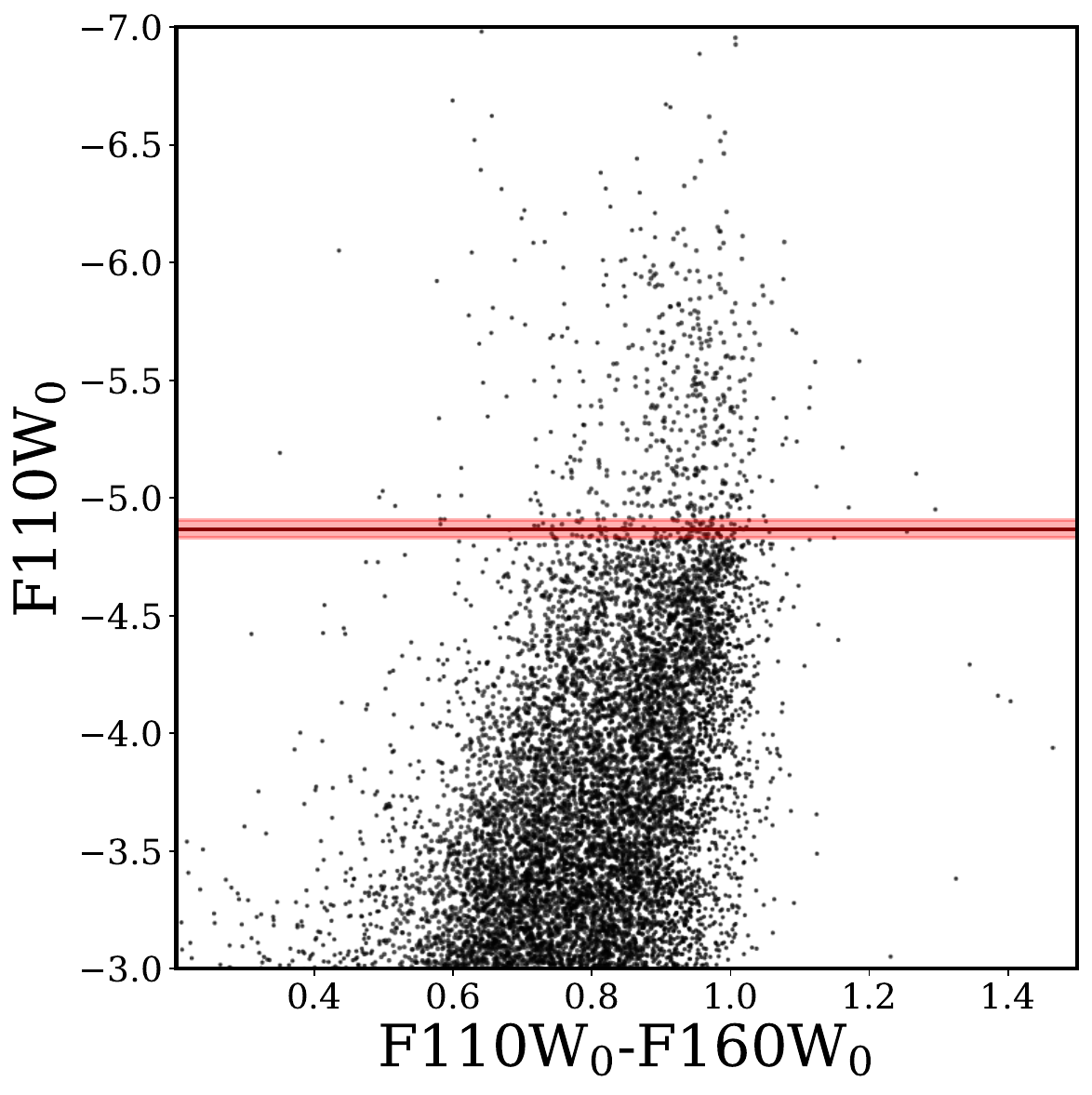}
    \includegraphics[width=0.35\textwidth]{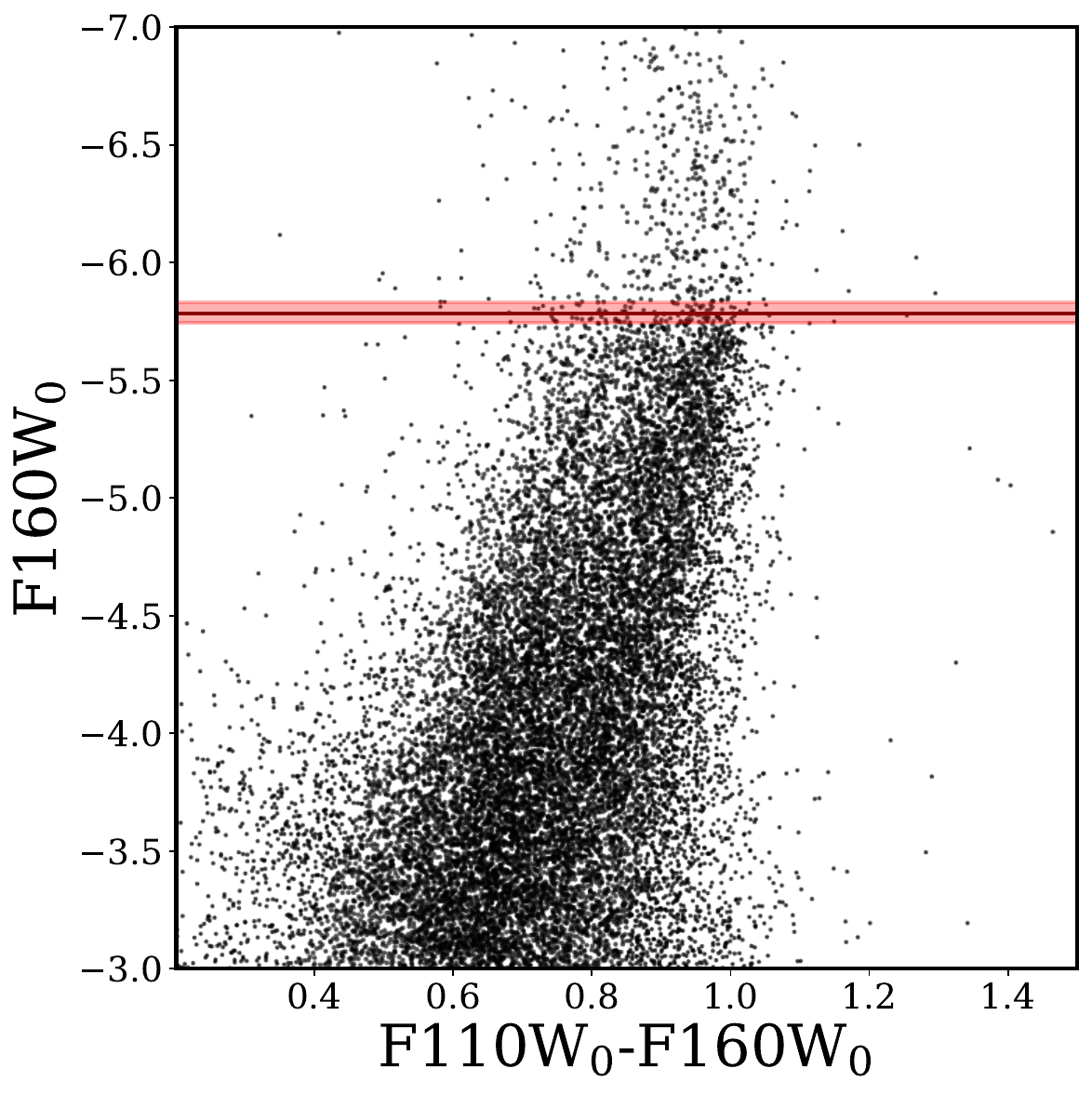} 
    \includegraphics[width=0.36\textwidth]{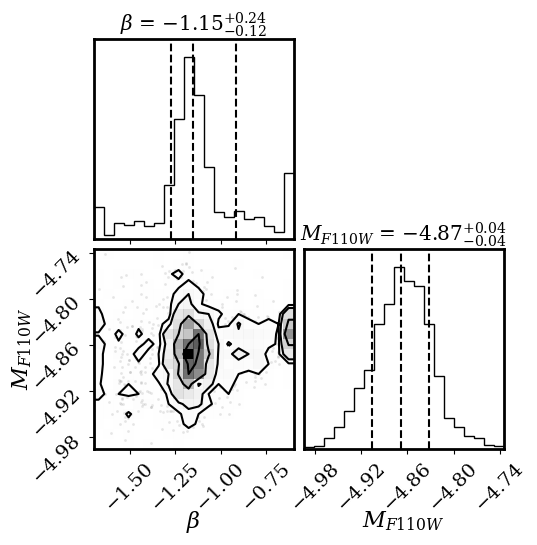}
    \includegraphics[width=0.36\textwidth]{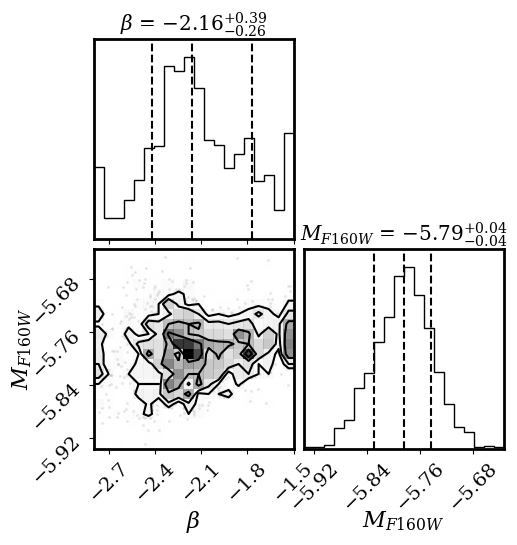}
	\caption{The final results from the NIR TRGB calibration MC simulations. The left and right sides of the figure show the results for F110W and F160W, respectively. (Top and middle panels) NIR CMDs with the best-fit calibration values (black solid lines) and uncertainties (red shaded region) overplotted. The CMDs include photometry from all galaxies after placing the catalogs on an absolute magnitude scale. The top two panels panels present the unrectified CMDs with slope fits of $\beta_{\text{F110W}}=\blueslope$~mag/mag and $\beta_{\text{F160W}}=\redslope$~mag/mag. The middle two panels present the rectified CMDs with zero point fits of $M^{WFC3}_{\text{F110W}}=\bluezp$~mag  and $M^{WFC3}_{\text{F160W}}=\redzp$~mag . Uncertainties are measured through 1500 MC simulations by simultaneously resampling the statistical uncertainties from $\mu_{\text{F814W}}$ and the photometric errors based on ASTs (see Section~\ref{sec:slope_and_zp_measure} for more details). (Bottom panel) The join probability distributions of the MC simulations for the slope and zero point fits. The $16^{\text{th}}, 50^{\text{th}}$ and $84^{\text{th}}$ percentile fits to the 1d histograms are shown as black dashed lines. The corner plot demonstrates a modest correlation between the slope and zero point fits and is propagated to the statistical uncertainties on the zero point.} 
	\label{fig:slope_zp_best_corner_w_uncertainties}
\end{figure*}
Figure~\ref{fig:slope_zp_best_corner_w_uncertainties} presents the resulting best-fit slopes (top), zero points (middle), and corner plots (bottom) with uncertainties on the best-fit values from the MC resampling procedure overplotted on the stacked F110W (left) and F160W (right) CMDs. The best-fit slopes and zero points are indicated by black solid lines and statistical uncertainties from the MC simulations are shown as red shaded regions about the fits. The CMDs in the top panels are the original, unrectified data to demonstrate how the luminosity of the TRGB depends on the color of the stars, while the CMDs in the middle two panels have been rectified using the best-fit slopes. The corner plots show the joint probability of the fits from the 1500 MC trials for each filter and the $16^{\text{th}}, 50^{\text{th}}$, and $84^{\text{th}}$ percentile fits to the 1d histograms as black dashed lines. We find that there is a modest correlation between the slopes and zero points in both filters. The slope and zero point values are shown in the bottom panels and are described in Eqs.~\ref{eqn:f110w_color_correction} and \ref{eqn:f160w_color_correction}. We again emphasize that these zero points are tied to the TRGB color of F110W-F160W $= 0.92$~mag.

Finally, we use the mathematical transformations derived in \S~\ref{sec:calib_math_basis} (Eqs.~\ref{eqn:transform_eq_zp} \& \ref{eqn:transform_eq_slope}) to check the empirically derived calibrations. In an ideal scenario it is guaranteed that the mathematical transformation and the empirically derived slopes and zero points are identical. However, if the slopes and zero points are not well fit, they can differ from each other. Our empirically fit $M^{\text{TRGB}}_{\text{F160W},\gamma}$ agrees exactly with the transformed zero point, while the empirically measured slope differs from the transformed slope by 0.01~mag/mag which is less than the uncertainty on the slope. Thus, the agreement is within uncertainties as expected.

\subsection{Error Budget}\label{sec:error_budget}
We provide a summary of our total error budget and tabulate the sources of uncertainty in Table~\ref{tab:error_budget}. The sources of uncertainty that we consider are:
\begin{itemize}
    \item Foreground extinction (sys)
    \item Smoothing Scale (sys)
    \item F814W QT color correction (sys)
    \item NIR Zero Point (stat)
    \item F814W Zero Point (stat and sys)
\end{itemize}
For the foreground extinction systematic uncertainty we adopt a systemic uncertainty of 0.01~mag equal to 10\% of the maximum extinction value in F814W. Based on the choice of GLOESS smoothing scale $\tau$ we determined a systematic uncertainty of 0.02~mag by re-running the MC simulation with smoothing scales of $\tau=0.02$ and $\tau=0.08$. We adopt an estimate of 0.01~mag systematic uncertainty for applying the F814W color correction applied to measure the TRGB in F814W. 
The statistical uncertainty on the NIR zero  points is $0.04$~mag in both filters.

Finally, we adopt the statistical uncertainty of $0.015$~mag and the systematic uncertainty of $0.035$~mag reported in \citet{Freedman2021} for the F814W TRGB zero point adopted in anchoring distance scale for our NIR TRGB calibration. As discussed in \S\ref{sec:acs_distances}, their systematic uncertainty is based on a weighted mean of F814W TRGB zero point calibrations tied to geometric anchors in the LMC, SMC, Galactic Globular Clusters, and the megamaser galaxy NGC~4258. The geometric distances to the LMC and SMC are determined from detached eclipsing binaries \citep{Pietrzynski2013, Pietrzynski2019, Graczyk2020}. The geometric distance to NGC~4258 is determined from its nuclear water megamaser sources \citep{Humphreys2013, Riess2016}. From Table~\ref{tab:error_budget}, we see that the dominant contribution to the statistical uncertainty is the NIR zero point and the dominant contribution to the systematic uncertainty is the F814W magnitude of the TRGB.

\begin{table}[!tbh]
    \centering
    \footnotesize
    \caption{Summary of the Error Budget for the NIR TRGB Calibration}
    \label{tab:error_budget}
    \begin{tabular}{lll}
    \hline
    \hline
    Source & Stat. (mag) & Sys. (mag) \\
    \hline
    Foreground Extinction & $\ldots$ & $10\% \text{max}(A_{\text{F814W}})=0.01$ \\
    Smoothing Scale & $\ldots$ & 0.02\\
    QT Color Correction\footnote{Only 2 of the 12 targets contribute to this systematic.} & $\ldots$ & 0.01\\
    $M_{\text{F814W}}^{TRGB}$ & 0.015 & 0.035\\
    NIR zero point & $0.04$ & $\ldots$\\
    \hline
    \multicolumn{3}{c}{\bf{Totals}}\\
    \hline
    & $0.043$ & $0.043$\\
    \hline
    \end{tabular}
\end{table}

\subsection{Comparison of F814W and NIR Distances}
In order to validate our NIR TRGB calibration, we re-measure the distances to each field in the individual NIR CMDs using our new calibration and compare them to the F814W distances. To measure the TRGB magnitudes in the F110W and F160W filters, we apply the same method used in the original F814W TRGB measurements. First, we apply spatial cuts to the photometry catalogs as described in \S\ref{sec:spatial_cuts} in order to reduce contamination from non-RGB populations. Second, we apply our color-based correction to each star within the color range $0.73\leq\text{F110W-F160W}\leq1.0$~mag  with the appropriate equation for F110W (see Eq.~\ref{eqn:f110w_color_correction}) or F160W (see Eq.~\ref{eqn:f160w_color_correction}). Third, we use the ML technique described in \S~\ref{sec:ACS_TRGB} to the foreground-extinction corrected NIR photometry. 

\begin{figure*}[!bht]
	\includegraphics[width=0.5\textwidth, height=0.47\textheight]{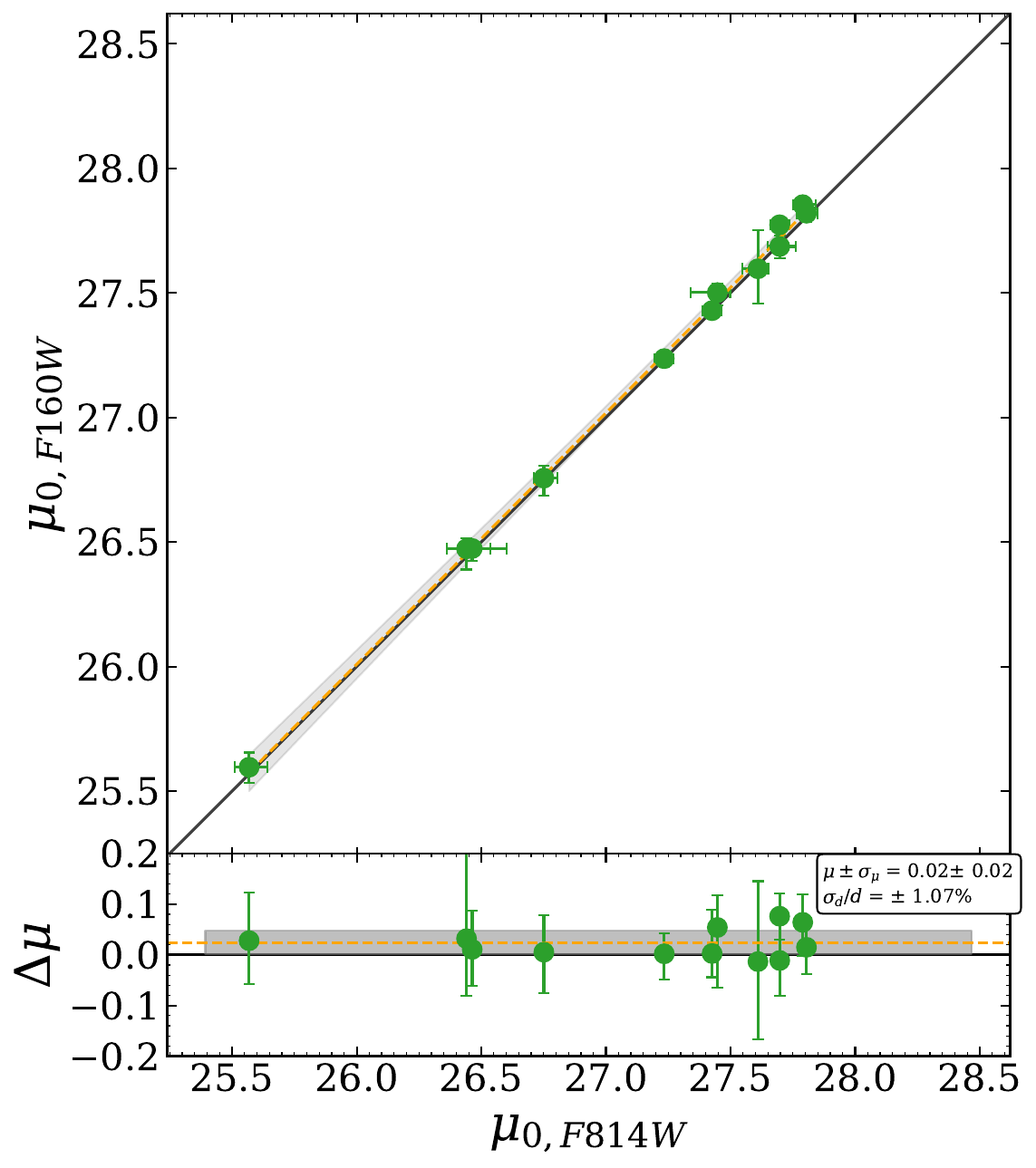}
	\includegraphics[width=0.5\textwidth, height=0.47\textheight]{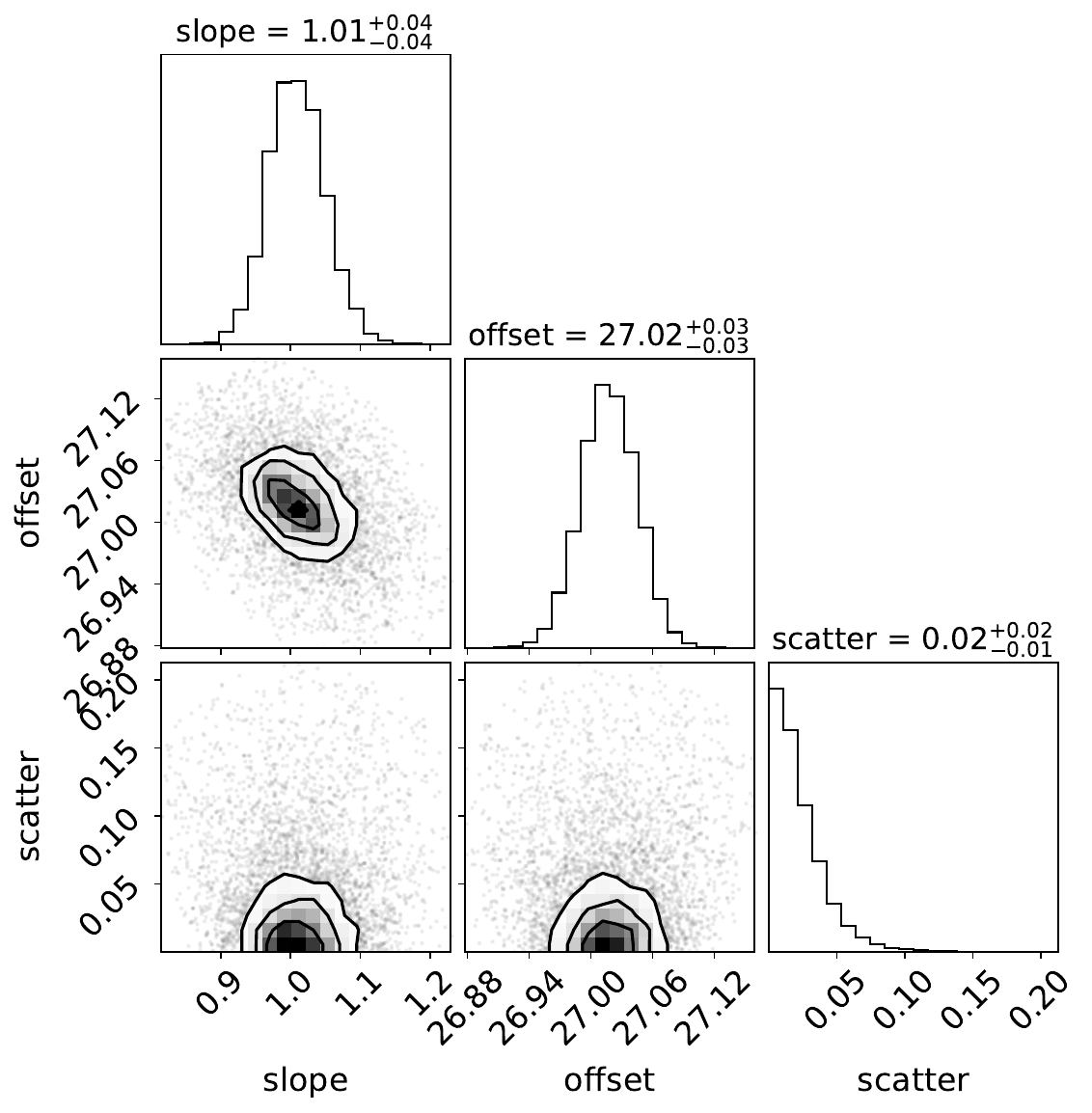}
	\caption{We present a comparison of the distance moduli based on the F814W TRGB and NIR F160W TRGB. { Left:} The top panel shows the distance modulus comparison (filled green circles). The orange dashed line and grey shaded region shows the linear regression fit to the green points, and the solid black line is the 1-to-1 line. We fit the data with \texttt{UltraNest} \citep{Buchner2021}. The bottom panel shows the residuals in DM $\Delta\mu = \mu_{0,\text{F160W}}-\mu_{0,\text{F814W}}$. The positive residuals indicate that the NIR modulus is fainter, thus the measured distance is further away, than the F814W-based distance. The black dashed line marks $\Delta\mu=0$, and the orange dashed line and gray shaded region are the mean and standard deviation of the mean weighted by the uncertainties on the data. The residual plot demonstrates that the distance moduli are in excellent agreement within the uncertainties based on both the weighted mean and standard deviation of the weighted mean of the residuals ($\bar{\mu}_{\text{F160W}}\approx-0.02\pm0.02$~(stat)). As a fractional uncertainty in distance, this represents a precision of $\sim1\%$. {Right:} Corner plot for the fit parameters (slope, offset, and intrinsic scatter) from the ML fit to the data in the distance moduli panel. The result of the fit is consistent with a 1-to-1 relationship between $\mu_{0, \text{F160W}}$ and $\mu_{0, \text{F814W}}$.}
	\label{fig:f160w_dm_comp}
\end{figure*}

We use the zero point to determine the distance moduli in the NIR filters and compare the results to the distance moduli as determined in the F814W filter. Figure~\ref{fig:f160w_dm_comp} presents the distance modulus comparison for the F160W and F814W results. The top panel shows the F160W distance moduli plotted as a function of the F814W distance moduli, with the top panel showing the distance moduli and the bottom panel showing the residuals between the distance moduli. We fit a linear model to the data and derive posterior probability distributions and the Bayesian
evidence with the nested sampling Monte Carlo algorithm
MLFriends (Buchner, 2014; 2019) using the
\texttt{UltraNest}\footnote{\url{https://johannesbuchner.github.io/UltraNest/}} package (Buchner 2021). The right side of each panel shows the corner plots the \texttt{UltraNest} regression fits to the orange points in the top of the distance modulus panel. Table~\ref{tab:table6} presents both the F110W and F160W TRGB fits and resulting distance moduli and distances. In both comparisons, the \texttt{UltraNest} fits to the distance moduli result in approximately 1-to-1 relationships (regression slope fit = 1.01$\pm0.04$; see the right panel in Figure \ref{fig:f160w_dm_comp}). Similarly, the standard deviation of the weighted mean based on distance modulus residuals is $\mu=0.02\pm0.02$, confirming that the distance moduli agree. The F160W and F814W distance moduli are in excellent agreement for every field indicating a very robust calibration.

\begin{table*}[!tbh]
	\centering
	\footnotesize
	\caption{Summary of Distances Measured from NIR LF}
	\label{tab:table6}
	\begin{tabular}{l | cccccc}
	\hline
	Galaxy  & m$^{\text{TRGB}}_{0,\text{F110W}}$   & $\mu_{0,\text{F110W}}$   & D$_{\text{F110W}}$   & m$^{\text{TRGB}}_{0,\text{F160W}}$  & $\mu_{0,\text{F160W}}$ & D$_{\text{F160W}}$\\ 
			& (mag)                    & (mag)		 & (Mpc)           & (mag) & (mag) & (Mpc)\\
	\hline
	\hline
    M81-1	& $22.98^{+0.02}_{-0.06}$	& $27.85^{+0.04}_{-0.07}$	& $3.71^{+0.07}_{-0.12}$	& $22.06^{+0.02}_{-0.05}$	& $27.85^{+0.07}_{-0.07}$	& $3.72^{+0.08}_{-0.11}$ \\
M81-2	& $22.94^{+0.13}_{-0.08}$	& $27.81^{+0.14}_{-0.09}$	& $3.64^{+0.23}_{-0.15}$	& $22.03^{+0.04}_{-0.04}$	& $27.82^{+0.05}_{-0.05}$	& $3.66^{+0.09}_{-0.09}$ \\
NGC253-1	& $22.85^{+0.02}_{-0.04}$	& $27.72^{+0.04}_{-0.06}$	& $3.50^{+0.07}_{-0.09}$	& $21.90^{+0.04}_{-0.05}$	& $27.69^{+0.06}_{-0.06}$	& $3.45^{+0.10}_{-0.10}$ \\
NGC253-2	& $22.91^{+0.02}_{-0.02}$	& $27.78^{+0.04}_{-0.04}$	& $3.60^{+0.07}_{-0.07}$	& $21.98^{+0.02}_{-0.03}$	& $27.77^{+0.05}_{-0.05}$	& $3.59^{+0.08}_{-0.08}$ \\
NGC2403-1	& $22.64^{+0.03}_{-0.05}$	& $27.51^{+0.05}_{-0.06}$	& $3.17^{+0.07}_{-0.09}$	& $21.71^{+0.04}_{-0.05}$	& $27.50^{+0.07}_{-0.07}$	& $3.16^{+0.08}_{-0.09}$ \\
NGC2403-2	& $22.56^{+0.09}_{-0.02}$	& $27.43^{+0.10}_{-0.04}$	& $3.07^{+0.14}_{-0.06}$	& $21.64^{+0.08}_{-0.03}$	& $27.43^{+0.05}_{-0.05}$	& $3.06^{+0.12}_{-0.07}$ \\
NGC300-1	& $21.60^{+0.04}_{-0.07}$	& $26.47^{+0.06}_{-0.08}$	& $1.97^{+0.05}_{-0.07}$	& $20.68^{+0.03}_{-0.05}$	& $26.47^{+0.07}_{-0.07}$	& $1.97^{+0.04}_{-0.06}$ \\
NGC300-2	& $21.62^{+0.05}_{-0.09}$	& $26.49^{+0.06}_{-0.10}$	& $1.98^{+0.06}_{-0.09}$	& $20.68^{+0.04}_{-0.08}$	& $26.47^{+0.09}_{-0.09}$	& $1.97^{+0.05}_{-0.08}$ \\
Antlia Dwarf	& $20.73^{+0.05}_{-0.06}$	& $25.60^{+0.07}_{-0.07}$	& $1.32^{+0.04}_{-0.04}$	& $19.81^{+0.06}_{-0.06}$	& $25.60^{+0.07}_{-0.07}$	& $1.32^{+0.04}_{-0.04}$ \\
UGC9128	& $21.89^{+0.06}_{-0.08}$	& $26.76^{+0.07}_{-0.09}$	& $2.24^{+0.07}_{-0.09}$	& $20.97^{+0.05}_{-0.07}$	& $26.76^{+0.08}_{-0.08}$	& $2.25^{+0.07}_{-0.08}$ \\
UGC8833	& $22.72^{+0.04}_{-0.03}$	& $27.59^{+0.06}_{-0.05}$	& $3.29^{+0.09}_{-0.08}$	& $21.81^{+0.15}_{-0.14}$	& $27.60^{+0.15}_{-0.15}$	& $3.31^{+0.24}_{-0.22}$ \\
UGC9240	& $22.35^{+0.03}_{-0.02}$	& $27.22^{+0.05}_{-0.05}$	& $2.78^{+0.06}_{-0.06}$	& $21.45^{+0.02}_{-0.03}$	& $27.24^{+0.05}_{-0.05}$	& $2.80^{+0.06}_{-0.07}$ \\
	\hline \hline
	\multicolumn{7}{l}{
		\begin{minipage}{12cm}~\\
			NOTE - We apply the color-based calibrations and adopt the F110W and F160W TRGB zero points from this study to determine the TRGB magnitude and distance modulus, respectively.
		\end{minipage}
	}
	\end{tabular}
\end{table*}

\section{Comparison with previous NIR TRGB Calibrations}\label{sec:discussion}
\begin{figure*}[!ht]
	\centering	\includegraphics[width=1\textwidth]{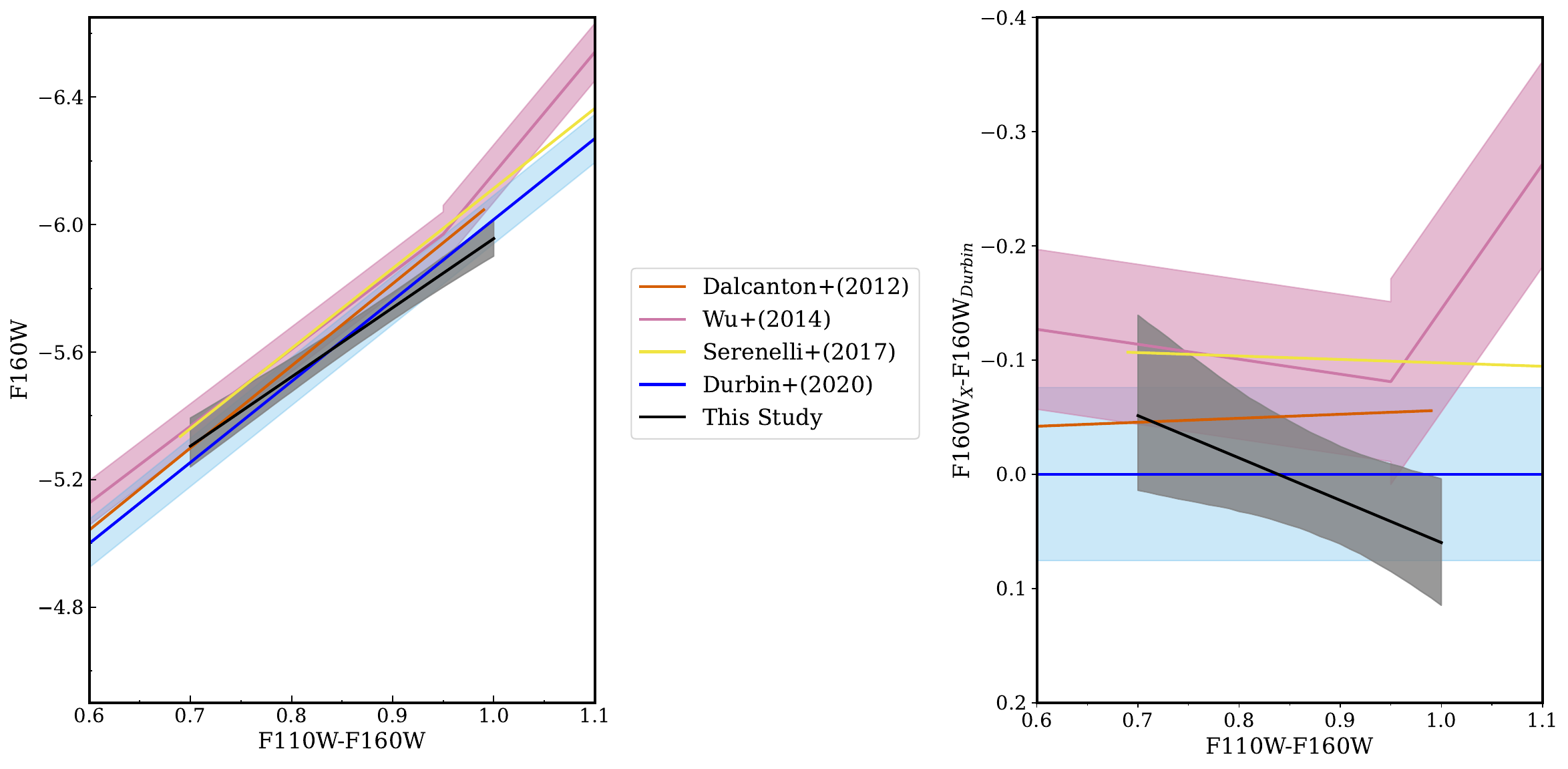}
	\caption{F160W TRGB color-magnitude relationships. Left: Calibrations from this work (gray) as well as other studies from the literature with uncertainties where available. Note that the uncertainties about the \citet{Wu2014} calibration are shown only for the zero point. The mismatch in the \citet{Wu2014} calibration where the slope changes at a color F110W-F160W$=0.95$ is due to differences in the uncertainties as a function of color. Right: The same color-magnitude relationships in the left panel now presented relative to the \citet{Durbin2020} calibration. The calibrations agree within the uncertainties over a narrow range in color with the \citet{Wu2014} calibration deviating the most significantly at the reddest colors.}
\label{fig:color_based_corr_comp}
\end{figure*}

Calibrations for the NIR TRGB have been reported in a number of studies based on HST photometry in WFC3/IR F110W and F160W, as well as based on theoretical stellar evolution models. These calibrations include color-based corrections, zero points, or both. In Figure~\ref{fig:color_based_corr_comp}, we compare the results of our F160W TRGB calibration with those from previous studies and provide a summary of the calibrations in Table~\ref{tab:table7}. Each calibration is plotted within the applicable color ranges (see Table~\ref{tab:table7} for the color ranges explored in each study). The individual calibrations are color-coded by the study from which they originate. The figure demonstrates the our calibration is in excellent agreement with the calibrations from the literature. We discuss individual differences between calibrations in greater detail below.

\citet{Dalcanton2012a} used data from 23 nearby galaxies ($D\lesssim4$~Mpc) to determine a calibration from the TRGB only in the F160W filter. They anchored their calibration to TRGB distances that they measured in the F814W filter calibrated from theoretical PARSEC isochrones \citep{Girardi2010, Bressan2012,Marigo2017}. To measure the NIR TRGB, \citet{Dalcanton2012a} applied the edge-detection filter described in \citet{Mendez2002} to a Gaussian-smoothed luminosity function \citep{Sakai1997} to each field. They first measured the apparent magnitude in each of their fields, then converted to absolute magnitudes by adopting their F814W TRGB-based distances as truth. However, there are known systematic differences in the predicated TRGB absolute magnitude depending on the adopted theoretical stellar evolutionary library which change both the shape and location of the TRGB as a function of color \cite[e.g.,][]{Beaton2018, Durbin2020}. Additionally, \citet{Dalcanton2012a} reported concerns that their analysis was impacted by the processing and calibration of WFC3/IR images, which at the time was in its early stages and has since improved. 

\citet{Durbin2020} re-reduced the data from \citet{Dalcanton2012a}, including updating the anchoring F814W TRGB-based distances, and applied a mutliwavelength-covariance method for measuring the TRGB across several HST filters. \citet{Durbin2020} leverage a combination of observed photometry and synthetic photometry to determine their multiwavelength TRGB measurements. However, they report a systematic uncertainty on the order of $\sim0.1$~mag depending on the adopted stellar evolutionary library. They caution a conservative 10\% systematic uncertainty included in any model-based absolute TRGB calibration. 

\cite{Wu2014} based their calibration on HST WFC3/IR data in the F110W and F160W filters for 32 fields; they combined the 23 galaxies covering 26 fields from \cite{Dalcanton2012a} with several M31 fields from the Panchromatic Hubble Andromeda Treasury \cite[PHAT;][]{Dalcanton2012b}. They adopted ACS/WFC F814W filter TRGB-based distance moduli from the Extragalactic Distance Database \cite[EDD; ][]{Tully2009} as their anchoring distances for all galaxies excluding M31. The EDD uses the extinction maps of \cite{Schlafly2011}, which are not well calibrated around M31, to determine reddening corrections. \citet{Wu2014} used the F814W TRGB magnitude before reddening corrections in combination with the extinction map from \citet{Montalto2009} to derive their own M31 distance modulus. They found a break in the TRGB magnitude color dependence at $\text{F110W-F160W}\approx0.95$. The change in slope at redder colors is predicted by stellar libraries and may be partially due to line-blanketing in the atmospheres of RGB stars where the metals preferentially absorb light at optical wavelengths and re-emit at NIR wavelengths. The line-blanketing effect is non-linear and may have a stronger effect at higher stellar metallicities leading to a non-linear increase in the color dependence of the TRGB luminosity function \citep{Barker2004, Mager2008, Wu2014}.

We find hints of a potential change in the slope in both F110W and F160W, albeit at a slightly bluer color than \citet{Wu2014}. However, our data are not suited to find the slope change to a high degree of confidence, and we therefore assume a linear dependence across our full color range. In addition, our calibration does not extend to the colors and magnitudes for M31 where \citet{Wu2014} find that the TRGB magnitude in the NIR may no longer be linear with color \citep{Barker2004, Mager2008}. Future data sets which include more stars across a wider range of colors can help constrain this effect. An empirical calibration of the slope of the TRGB at NIR wavelengths that has a wider metallicity range has wide applications to both observers and to stellar modelers.

Over the color range in our calibration, the method that \cite{Wu2014} used to calibrate the TRGB in the F110W and F160W filters differs from the method developed in this study. They use the ML method described in Section~\ref{sec:acs_distances} to measure the TRGB in each WFC3/IR filter prior to correcting for any color dependence, and derive a TRGB color from the median color of stars within a magnitude interval of 0.05 mag fainter than the measured TRGB. In contrast, we first determine our color-based correction from our combined photometry catalog, then measure the TRGB magnitudes from the rectified CMDs. These differences between our calibration and the \cite{Wu2014} calibration may help explain the offset as shown in Figure~\ref{fig:color_based_corr_comp}.

\citet{Serenelli2017} report a theoretical calibration for the NIR TRGB-based on two different stellar evolutionary codes, \texttt{BaSTI} \citep{Pietrinferni2004,Pietrinferni2006} and \texttt{GARSTEC} \citep{Weiss2008}. \citet{Serenelli2017} use bolometric corrections determined from theoretical and empirical calibrations to measure the color-magnitude relationship of the F110W and F160W TRGB. They report that uncertainties in the bolometric corrections make it challenging to generate an accurate reference set for TRGB absolute magnitudes. We find that our new calibration is in overall agreement with previous calibrations. 

\begin{table*}[!t]
    \centering
    \footnotesize
    \caption{Summary of Previous NIR TRGB Calibrations in the WFC3/IR F160W Filter}
    \label{tab:table7}
    \begin{tabular}{llc}
    \hline
    \hline
    Reference &  Calibration ($M_{\text{F160W}}^{\text{TRGB}}$)& Color Range\\
    \hline
    This Study & $\redzp\redslope\left[\left(\text{F110W-F160W}\right)-0.92\right]$ & $0.73\leq\text{F110W-F160W}\leq1.0$\\
    \cite{Dalcanton2012a} & $-2.567\left(\text{F110W-F160W}\right)-3.495$ & - \\ 
    \cite{Wu2014} & $-5.97-2.41\left[\left(\text{F110W-F160W}\right)-0.95\right]$ & $\text{F110W-F160W}\leq0.95$\\ 
                  & $-5.97-3.81\left[\left(\text{F110W-F160W}\right)-0.95\right]$  & $\text{F110W-F160W}>0.95$\\
     \cite{Serenelli2017} & $-5.310-2.511\left[\left(\text{F110W-F160W}\right)-0.68\right]$ & $0.68\leq\text{F110W-F160W}<1.20$\\
    \cite{Durbin2020} & $-2.541\left(\text{F110W-F160W}\right)-3.475$ & -\\
    \hline
    \hline
    \end{tabular}
\end{table*}

\section{Conclusion}\label{sec:summ_conc}
The TRGB in the NIR has the potential to expand the range over which efficient and precise distance measurements are possible. However, for the NIR TRGB to be useful as a precision distance indicator it must be carefully calibrated. We presented an entirely empirical calibration, relative to the F814W TRGB calibration, for the TRGB in the WFC3/IR F110W and F160W filters including color-based corrections and zero point calibrations accurate to $\pm0.043$~(stat) $\pm0.043$ (sys)~mag in distance modulus based on the standard deviation of the mean of the residuals weighted by the individual distance modulus uncertainties, shown in Figure~\ref{fig:f160w_dm_comp} as the gray shaded region in the residual plot. 

We derived high-fidelity cross-matched photometric catalogs based on observations of 8 galaxies and 12 fields in F606W and F814W filters, and WFC3/IR F110W and F160W filters. From the photometric catalogs we measured TRGB-based distances in the F814W filter using the empirical QT calibration from \citep{Jang2017} and zero point from \citep{Freedman2021} to anchor our calibration on a uniform distance scale.

To measure the color-magnitude relationship in the F110W and F160W filters we generated a combined stellar catalog from the entire sample to provide the broadest metallicity baseline. From the combined stellar catalog we measure the slope, color pivot point, and zero point of the TRGB in the F110W and F160W filters. We provide the results of our calibration with uncertainties in \S\ref{sec:slope_and_zp_measure} \& \S\ref{sec:error_budget}
(Eqs.~\ref{eqn:f110w_color_correction}~\& \ref{eqn:f160w_color_correction}). Finally we verified the internal precision of the calibration to $\sim1\%$ in $D$ based on a fit to the mean and weighted standard deviation of the mean in the distance modulus residuals $\mu_{\text{F160W}}- \mu_{\text{F814W}}$. These new calibrations enable distance measurements over a larger volume than is accessible with the ACS F814W TRGB method.

The calibration presented in this study is robustly measured in the WFC3/IR F110W and F160W filters within the color range $0.73\leq\text{F110W-F160W}\leq1.0$~mag. We emphasize that, similarly to in the F814W filter, the TRGB in the F110W and F160W filters is most easily identified when minimal non-RGB stellar populations are present in the CMD. These include AGB and RHeB stars that can mask the location of the TRGB in a CMD, especially since the color baseline of the RGB is narrow. Careful choice of observing fields can help mitigate the masking effect from non-RGB populations. In particular, imaging the outer stellar fields of galaxies may help to reduce the impact of non-RGB populations, since we expect older RGB stars to dominate in this spatial regime. The next phase of the NIR TRGB calibration is to measure color-based corrections and zero points in the JWST filters on the NIRCam and NIRISS instruments as part of JWST-GO-1638 (PI McQuinn). 

\facility{HST (ACS, WFC3)}
\software{Astropy \citep{AstropyCollaboration2022}, \Dolphot{} \citep{Dolphin2002,Dolphin2016}, corner \citep{corner}, Matplotlib \citep{Hunter:2007}, NumPy \citep{harris2020array}, SciPy \citep{2020SciPy-NMeth}, UltraNest \citep{Buchner2021}}

\acknowledgments
We thank the referee, whose insightful comments significantly improved the paper. Support for program HST-GO-15917 was provided by NASA through a grant from the Space Telescope Science Institute, which is operated by the Associations of Universities for Research in Astronomy, Incorporated, under NASA contract NAS 5-26555. This research was supported by the Munich Institute for Astro-, Particle and BioPhysics (MIAPbP), which is funded by the Deutsche Forschungsgemeinschaft (DFG, German Research Foundation) under Germany´s Excellence Strategy – EXC-2094 – 390783311. This research has made use of the NASA/IPAC Extragalactic Database (NED), which is funded by the National Aeronautics and Space Administration and operated by the California Institute of Technology. All the HST data used in this paper can be found in MAST at:\dataset[10.17909/faaj-4934]{http://dx.doi.org/10.17909/faaj-4934} and\dataset[10.17909/mr9z-6882]{http://dx.doi.org/10.17909/mr9z-6882}. 
This research has made use of adstex (\url{https://github.com/yymao/adstex}).

\appendix
\section{Full ACS/WFC FOV CMDs}
\begin{figure}
    \centering
    \includegraphics[width=\textwidth]{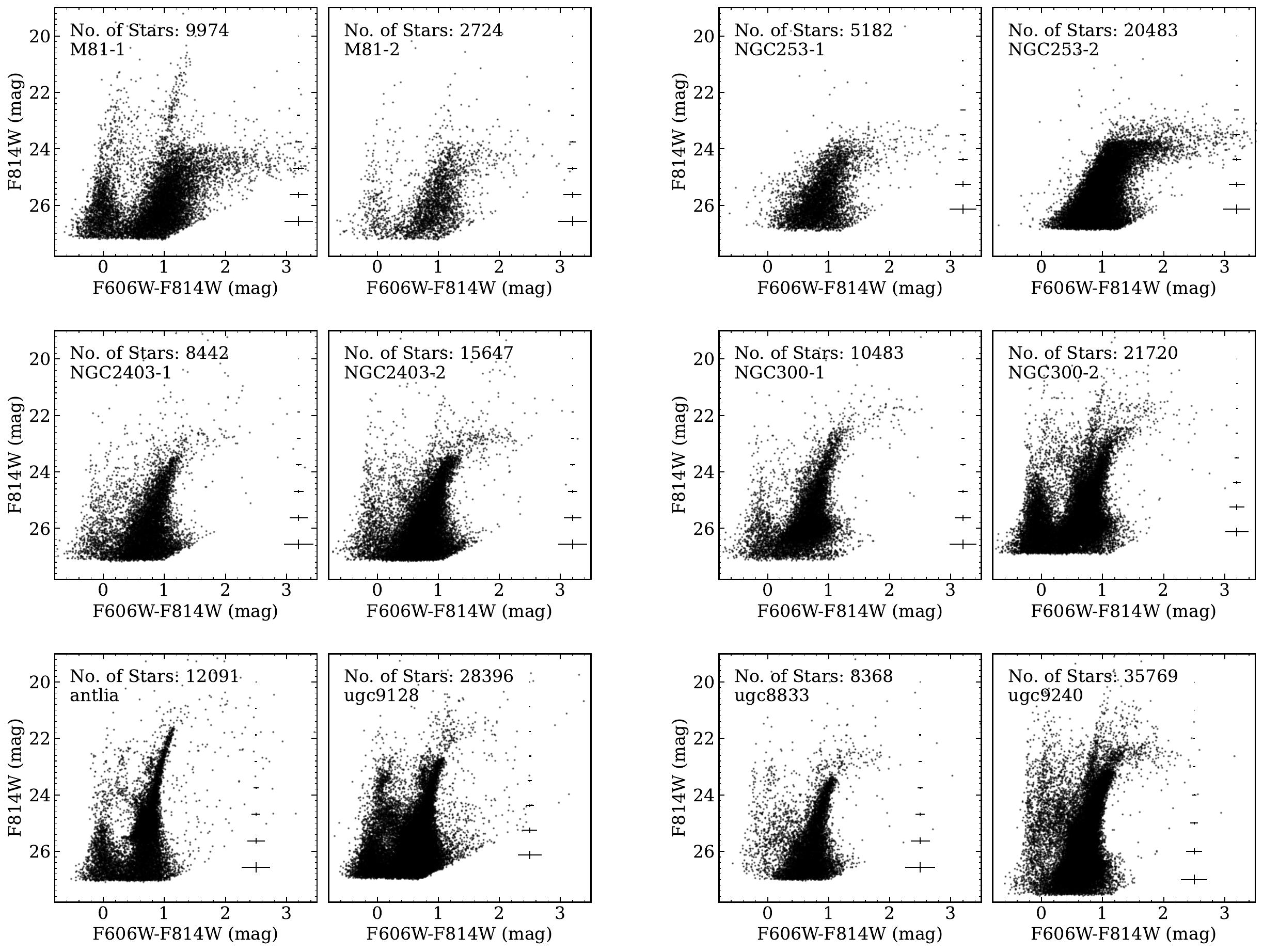}
    \caption{Full ACS/WFC optical CMDs. Counterclockwise from the top left, left to right in each pair: M81 (fields: 1,2), NGC~253 (fields: 1,2), NGC~300 (fields: 1,2), UGC~8833, UGC~9240, Antlia Dwarf, UGC9128, NGC~2403 (fields: 1,2). The number of stars in each CMD are shown in the legends.}
    \label{fig:fullACS_appendix}
\end{figure}

In Figure \ref{fig:fullACS_appendix} we present the optical CMDs from the full ACS/WFC FOV for the new observations (M81, NGC~253, NGC~300, NGC~2403) and for the archival observations (Antlia Dwarf, UGC~9128, UGC~8833, and UGC~9240).

\renewcommand\bibname{References}
\bibliography{ms.bib}

\begin{thebibliography}{}
\expandafter\ifx\csname natexlab\endcsname\relax\def\natexlab#1{#1}\fi
\providecommand{\url}[1]{\href{#1}{#1}}
\providecommand{\dodoi}[1]{doi:~\href{http://doi.org/#1}{\nolinkurl{#1}}}
\providecommand{\doeprint}[1]{\href{http://ascl.net/#1}{\nolinkurl{http://ascl.net/#1}}}
\providecommand{\doarXiv}[1]{\href{https://arxiv.org/abs/#1}{\nolinkurl{https://arxiv.org/abs/#1}}}

\bibitem[{{Ahumada} {et~al.}(2020){Ahumada}, {Allende Prieto}, {Almeida},
  {Anders}, {Anderson}, {Andrews}, {Anguiano}, {Arcodia}, {Armengaud},
  {Aubert}, \& et~al.}]{Ahumada+2020}
{Ahumada}, R., {Allende Prieto}, C., {Almeida}, A., {et~al.} 2020, \apjs, 249,
  3, \dodoi{10.3847/1538-4365/ab929e}

\bibitem[{{Alam} {et~al.}(2015){Alam}, {Albareti}, {Allende Prieto}, {Anders},
  {Anderson}, {Anderton}, {Andrews}, {Armengaud}, {Aubourg}, {Bailey}, \&
  et~al.}]{Alam2015}
{Alam}, S., {Albareti}, F.~D., {Allende Prieto}, C., {et~al.} 2015, \apjs, 219,
  12, \dodoi{10.1088/0067-0049/219/1/12}

\bibitem[{{Astropy Collaboration} {et~al.}(2022){Astropy Collaboration},
  {Price-Whelan}, {Lim}, {Earl}, {Starkman}, {Bradley}, {Shupe}, {Patil},
  {Corrales}, {Brasseur}, {N{\"o}the}, {Donath}, {Tollerud}, {Morris},
  {Ginsburg}, {Vaher}, {Weaver}, {Tocknell}, {Jamieson}, {van Kerkwijk},
  {Robitaille}, {Merry}, {Bachetti}, {G{\"u}nther}, {Aldcroft},
  {Alvarado-Montes}, {Archibald}, {B{\'o}di}, {Bapat}, {Barentsen},
  {Baz{\'a}n}, {Biswas}, {Boquien}, {Burke}, {Cara}, {Cara}, {Conroy},
  {Conseil}, {Craig}, {Cross}, {Cruz}, {D'Eugenio}, {Dencheva}, {Devillepoix},
  {Dietrich}, {Eigenbrot}, {Erben}, {Ferreira}, {Foreman-Mackey}, {Fox},
  {Freij}, {Garg}, {Geda}, {Glattly}, {Gondhalekar}, {Gordon}, {Grant},
  {Greenfield}, {Groener}, {Guest}, {Gurovich}, {Handberg}, {Hart},
  {Hatfield-Dodds}, {Homeier}, {Hosseinzadeh}, {Jenness}, {Jones}, {Joseph},
  {Kalmbach}, {Karamehmetoglu}, {Ka{\l}uszy{\'n}ski}, {Kelley}, {Kern},
  {Kerzendorf}, {Koch}, {Kulumani}, {Lee}, {Ly}, {Ma}, {MacBride}, {Maljaars},
  {Muna}, {Murphy}, {Norman}, {O'Steen}, {Oman}, {Pacifici}, {Pascual},
  {Pascual-Granado}, {Patil}, {Perren}, {Pickering}, {Rastogi}, {Roulston},
  {Ryan}, {Rykoff}, {Sabater}, {Sakurikar}, {Salgado}, {Sanghi}, {Saunders},
  {Savchenko}, {Schwardt}, {Seifert-Eckert}, {Shih}, {Jain}, {Shukla}, {Sick},
  {Simpson}, {Singanamalla}, {Singer}, {Singhal}, {Sinha}, {Sip{\H{o}}cz},
  {Spitler}, {Stansby}, {Streicher}, {{\v{S}}umak}, {Swinbank}, {Taranu},
  {Tewary}, {Tremblay}, {de Val-Borro}, {Van Kooten}, {Vasovi{\'c}}, {Verma},
  {de Miranda Cardoso}, {Williams}, {Wilson}, {Winkel}, {Wood-Vasey}, {Xue},
  {Yoachim}, {Zhang}, {Zonca}, \& {Astropy Project
  Contributors}}]{AstropyCollaboration2022}
{Astropy Collaboration}, {Price-Whelan}, A.~M., {Lim}, P.~L., {et~al.} 2022,
  \apj, 935, 167, \dodoi{10.3847/1538-4357/ac7c74}

\bibitem[{{Avila} {et~al.}(2015){Avila}, {Murray}, {Knebe}, {Power},
  {Robotham}, \& {Garcia-Bellido}}]{Avila2015}
{Avila}, S., {Murray}, S.~G., {Knebe}, A., {et~al.} 2015, \mnras, 450, 1856,
  \dodoi{10.1093/mnras/stv711}

\bibitem[{{Barker} {et~al.}(2004){Barker}, {Sarajedini}, \&
  {Harris}}]{Barker2004}
{Barker}, M.~K., {Sarajedini}, A., \& {Harris}, J. 2004, \apj, 606, 869,
  \dodoi{10.1086/383026}

\bibitem[{{Beaton} {et~al.}(2016){Beaton}, {Freedman}, {Madore}, {Bono},
  {Carlson}, {Clementini}, {Durbin}, {Garofalo}, {Hatt}, {Jang}, {Kollmeier},
  {Lee}, {Monson}, {Rich}, {Scowcroft}, {Seibert}, {Sturch}, \&
  {Yang}}]{Beaton2016}
{Beaton}, R.~L., {Freedman}, W.~L., {Madore}, B.~F., {et~al.} 2016, \apj, 832,
  210, \dodoi{10.3847/0004-637X/832/2/210}

\bibitem[{{Beaton} {et~al.}(2018){Beaton}, {Bono}, {Braga}, {Dall'Ora},
  {Fiorentino}, {Jang}, {Mart{\'\i}nez-V{\'a}zquez}, {Matsunaga}, {Monelli},
  {Neeley}, \& {Salaris}}]{Beaton2018}
{Beaton}, R.~L., {Bono}, G., {Braga}, V.~F., {et~al.} 2018, \ssr, 214, 113,
  \dodoi{10.1007/s11214-018-0542-1}

\bibitem[{{Bressan} {et~al.}(2012){Bressan}, {Marigo}, {Girardi}, {Salasnich},
  {Dal Cero}, {Rubele}, \& {Nanni}}]{Bressan2012}
{Bressan}, A., {Marigo}, P., {Girardi}, L., {et~al.} 2012, \mnras, 427, 127,
  \dodoi{10.1111/j.1365-2966.2012.21948.x}

\bibitem[{{Buchner}(2021)}]{Buchner2021}
{Buchner}, J. 2021, The Journal of Open Source Software, 6, 3001,
  \dodoi{10.21105/joss.03001}

\bibitem[{{Carretta} {et~al.}(2000){Carretta}, {Gratton}, {Clementini}, \&
  {Fusi Pecci}}]{Carretta2000}
{Carretta}, E., {Gratton}, R.~G., {Clementini}, G., \& {Fusi Pecci}, F. 2000,
  \apj, 533, 215, \dodoi{10.1086/308629}

\bibitem[{{Cohen} {et~al.}(2020){Cohen}, {Goudfrooij}, {Correnti}, {Gnedin},
  {Harris}, {Chandar}, {Puzia}, \& {S{\'a}nchez-Janssen}}]{Cohen2020}
{Cohen}, R.~E., {Goudfrooij}, P., {Correnti}, M., {et~al.} 2020, \apj, 890, 52,
  \dodoi{10.3847/1538-4357/ab64e9}

\bibitem[{{Conroy} {et~al.}(2019){Conroy}, {Naidu}, {Zaritsky}, {Bonaca},
  {Cargile}, {Johnson}, \& {Caldwell}}]{Conroy2019}
{Conroy}, C., {Naidu}, R.~P., {Zaritsky}, D., {et~al.} 2019, \apj, 887, 237,
  \dodoi{10.3847/1538-4357/ab5710}

\bibitem[{{Da Costa} \& {Armandroff}(1990)}]{DaCosta1990}
{Da Costa}, G.~S., \& {Armandroff}, T.~E. 1990, \aj, 100, 162,
  \dodoi{10.1086/115500}

\bibitem[{{Dalcanton} {et~al.}(2009){Dalcanton}, {Williams}, {Seth}, {Dolphin},
  {Holtzman}, {Rosema}, {Skillman}, {Cole}, {Girardi}, {Gogarten},
  {Karachentsev}, {Olsen}, {Weisz}, {Christensen}, {Freeman}, {Gilbert},
  {Gallart}, {Harris}, {Hodge}, {de Jong}, {Karachentseva}, {Mateo}, {Stetson},
  {Tavarez}, {Zaritsky}, {Governato}, \& {Quinn}}]{Dalcanton2009}
{Dalcanton}, J.~J., {Williams}, B.~F., {Seth}, A.~C., {et~al.} 2009, \apjs,
  183, 67, \dodoi{10.1088/0067-0049/183/1/67}

\bibitem[{{Dalcanton} {et~al.}(2012a){Dalcanton}, {Williams}, {Melbourne},
  {Girardi}, {Dolphin}, {Rosenfield}, {Boyer}, {de Jong}, {Gilbert}, {Marigo},
  {Olsen}, {Seth}, \& {Skillman}}]{Dalcanton2012a}
{Dalcanton}, J.~J., {Williams}, B.~F., {Melbourne}, J.~L., {et~al.} 2012a,
  \apjs, 198, 6, \dodoi{10.1088/0067-0049/198/1/6}

\bibitem[{{Dalcanton} {et~al.}(2012b){Dalcanton}, {Williams}, {Lang}, {Lauer},
  {Kalirai}, {Seth}, {Dolphin}, {Rosenfield}, {Weisz}, {Bell}, {Bianchi},
  {Boyer}, {Caldwell}, {Dong}, {Dorman}, {Gilbert}, {Girardi}, {Gogarten},
  {Gordon}, {Guhathakurta}, {Hodge}, {Holtzman}, {Johnson}, {Larsen}, {Lewis},
  {Melbourne}, {Olsen}, {Rix}, {Rosema}, {Saha}, {Sarajedini}, {Skillman}, \&
  {Stanek}}]{Dalcanton2012b}
{Dalcanton}, J.~J., {Williams}, B.~F., {Lang}, D., {et~al.} 2012b, \apjs, 200,
  18, \dodoi{10.1088/0067-0049/200/2/18}

\bibitem[{{Dolphin}(2016)}]{Dolphin2016}
{Dolphin}, A. 2016, {DOLPHOT: Stellar photometry}, Astrophysics Source Code
  Library, record ascl:1608.013.
\newblock \doeprint{1608.013}

\bibitem[{{Dolphin}(2000)}]{Dolphin2000}
{Dolphin}, A.~E. 2000, \pasp, 112, 1383, \dodoi{10.1086/316630}

\bibitem[{{Dolphin}(2002)}]{Dolphin2002}
{Dolphin}, A.~E. 2002, \mnras, 332, 91,
  \dodoi{10.1046/j.1365-8711.2002.05271.x}

\bibitem[{{Durbin} {et~al.}(2020){Durbin}, {Beaton}, {Dalcanton}, {Williams},
  \& {Boyer}}]{Durbin2020}
{Durbin}, M.~J., {Beaton}, R.~L., {Dalcanton}, J.~J., {Williams}, B.~F., \&
  {Boyer}, M.~L. 2020, \apj, 898, 57, \dodoi{10.3847/1538-4357/ab9cbb}

\bibitem[{{Ferraro} {et~al.}(2000){Ferraro}, {Montegriffo}, {Origlia}, \& {Fusi
  Pecci}}]{Ferraro2000}
{Ferraro}, F.~R., {Montegriffo}, P., {Origlia}, L., \& {Fusi Pecci}, F. 2000,
  \aj, 119, 1282, \dodoi{10.1086/301269}

\bibitem[{{Ford} {et~al.}(1998){Ford}, {Bartko}, {Bely}, {Broadhurst},
  {Burrows}, {Cheng}, {Clampin}, {Crocker}, {Feldman}, {Golimowski}, {Hartig},
  {Illingworth}, {Kimble}, {Lesser}, {Miley}, {Neff}, {Postman}, {Sparks},
  {Tsvetanov}, {White}, {Sullivan}, {Krebs}, {Leviton}, {La Jeunesse},
  {Burmester}, {Fike}, {Johnson}, {Slusher}, {Volmer}, \&
  {Woodruff}}]{Ford1998}
{Ford}, H.~C., {Bartko}, F., {Bely}, P.~Y., {et~al.} 1998, in Society of
  Photo-Optical Instrumentation Engineers (SPIE) Conference Series, Vol. 3356,
  Space Telescopes and Instruments V, ed. P.~Y. {Bely} \& J.~B. {Breckinridge},
  234--248, \dodoi{10.1117/12.324464}

\bibitem[{{Foreman-Mackey}(2016)}]{corner}
{Foreman-Mackey}, D. 2016, The Journal of Open Source Software, 1, 24,
  \dodoi{10.21105/joss.00024}

\bibitem[{{Freedman}(1988)}]{Freedman1988}
{Freedman}, W.~L. 1988, \aj, 96, 1248, \dodoi{10.1086/114878}

\bibitem[{{Freedman}(2021)}]{Freedman2021}
{Freedman}, W.~L. 2021, \apj, 919, 16, \dodoi{10.3847/1538-4357/ac0e95}

\bibitem[{{Freedman} {et~al.}(2019){Freedman}, {Madore}, {Hatt}, {Hoyt},
  {Jang}, {Beaton}, {Burns}, {Lee}, {Monson}, {Neeley}, {Phillips}, {Rich}, \&
  {Seibert}}]{Freedman2019}
{Freedman}, W.~L., {Madore}, B.~F., {Hatt}, D., {et~al.} 2019, \apj, 882, 34,
  \dodoi{10.3847/1538-4357/ab2f73}

\bibitem[{{Freedman} {et~al.}(2020){Freedman}, {Madore}, {Hoyt}, {Jang},
  {Beaton}, {Lee}, {Monson}, {Neeley}, \& {Rich}}]{Freedman2020}
{Freedman}, W.~L., {Madore}, B.~F., {Hoyt}, T., {et~al.} 2020, \apj, 891, 57,
  \dodoi{10.3847/1538-4357/ab7339}

\bibitem[{{Gilbert} {et~al.}(2014){Gilbert}, {Kalirai}, {Guhathakurta},
  {Beaton}, {Geha}, {Kirby}, {Majewski}, {Patterson}, {Tollerud}, {Bullock},
  {Tanaka}, \& {Chiba}}]{Gilbert2014}
{Gilbert}, K.~M., {Kalirai}, J.~S., {Guhathakurta}, P., {et~al.} 2014, \apj,
  796, 76, \dodoi{10.1088/0004-637X/796/2/76}

\bibitem[{{Girardi} {et~al.}(2010){Girardi}, {Williams}, {Gilbert},
  {Rosenfield}, {Dalcanton}, {Marigo}, {Boyer}, {Dolphin}, {Weisz},
  {Melbourne}, {Olsen}, {Seth}, \& {Skillman}}]{Girardi2010}
{Girardi}, L., {Williams}, B.~F., {Gilbert}, K.~M., {et~al.} 2010, \apj, 724,
  1030, \dodoi{10.1088/0004-637X/724/2/1030}

\bibitem[{{Gordon} {et~al.}(2016){Gordon}, {Fouesneau}, {Arab}, {Tchernyshyov},
  {Weisz}, {Dalcanton}, {Williams}, {Bell}, {Bianchi}, {Boyer}, {Choi},
  {Dolphin}, {Girardi}, {Hogg}, {Kalirai}, {Kapala}, {Lewis}, {Rix},
  {Sandstrom}, \& {Skillman}}]{Gordon2016}
{Gordon}, K.~D., {Fouesneau}, M., {Arab}, H., {et~al.} 2016, \apj, 826, 104,
  \dodoi{10.3847/0004-637X/826/2/104}

\bibitem[{{G{\'o}rski} {et~al.}(2018){G{\'o}rski}, {Pietrzy{\'n}ski}, {Gieren},
  {Graczyk}, {Suchomska}, {Karczmarek}, {Cohen}, {Zgirski}, {Wielg{\'o}rski},
  {Pilecki}, {Taormina}, {Ko{\l}aczkowski}, \& {Narloch}}]{Gorski2018}
{G{\'o}rski}, M., {Pietrzy{\'n}ski}, G., {Gieren}, W., {et~al.} 2018, \aj, 156,
  278, \dodoi{10.3847/1538-3881/aaeacb}

\bibitem[{{Graczyk} {et~al.}(2020){Graczyk}, {Pietrzy{\'n}ski}, {Thompson},
  {Gieren}, {Zgirski}, {Villanova}, {G{\'o}rski}, {Wielg{\'o}rski},
  {Karczmarek}, {Narloch}, {Pilecki}, {Taormina}, {Smolec}, {Suchomska},
  {Gallenne}, {Nardetto}, {Storm}, {Kudritzki}, {Ka{\l}uszy{\'n}ski}, \&
  {Pych}}]{Graczyk2020}
{Graczyk}, D., {Pietrzy{\'n}ski}, G., {Thompson}, I.~B., {et~al.} 2020, \apj,
  904, 13, \dodoi{10.3847/1538-4357/abbb2b}

\bibitem[{{Hack} {et~al.}(2013){Hack}, {Dencheva}, \& {Fruchter}}]{Hack2013}
{Hack}, W.~J., {Dencheva}, N., \& {Fruchter}, A.~S. 2013, in Astronomical
  Society of the Pacific Conference Series, Vol. 475, Astronomical Data
  Analysis Software and Systems XXII, ed. D.~N. {Friedel}, 49

\bibitem[{{Harris} {et~al.}(2020){Harris}, {Millman}, {van der Walt},
  {Gommers}, {Virtanen}, {Cournapeau}, {Wieser}, {Taylor}, {Berg}, {Smith},
  {Kern}, {Picus}, {Hoyer}, {van Kerkwijk}, {Brett}, {Haldane}, {del R{\'\i}o},
  {Wiebe}, {Peterson}, {G{\'e}rard-Marchant}, {Sheppard}, {Reddy}, {Weckesser},
  {Abbasi}, {Gohlke}, \& {Oliphant}}]{harris2020array}
{Harris}, C.~R., {Millman}, K.~J., {van der Walt}, S.~J., {et~al.} 2020, \nat,
  585, 357, \dodoi{10.1038/s41586-020-2649-2}

\bibitem[{{Hatt} {et~al.}(2017){Hatt}, {Beaton}, {Freedman}, {Madore}, {Jang},
  {Hoyt}, {Lee}, {Monson}, {Rich}, {Scowcroft}, \& {Seibert}}]{Hatt2017}
{Hatt}, D., {Beaton}, R.~L., {Freedman}, W.~L., {et~al.} 2017, \apj, 845, 146,
  \dodoi{10.3847/1538-4357/aa7f73}

\bibitem[{{Hoyt}(2021)}]{Hoyt2021}
{Hoyt}, T.~J. 2021, arXiv e-prints, arXiv:2106.13337,
  \dodoi{10.48550/arXiv.2106.13337}

\bibitem[{{Hoyt} {et~al.}(2018){Hoyt}, {Freedman}, {Madore}, {Seibert},
  {Beaton}, {Hatt}, {Jang}, {Lee}, {Monson}, \& {Rich}}]{Hoyt2018}
{Hoyt}, T.~J., {Freedman}, W.~L., {Madore}, B.~F., {et~al.} 2018, \apj, 858,
  12, \dodoi{10.3847/1538-4357/aab7ed}

\bibitem[{{Humphreys} {et~al.}(2013){Humphreys}, {Reid}, {Moran}, {Greenhill},
  \& {Argon}}]{Humphreys2013}
{Humphreys}, E.~M.~L., {Reid}, M.~J., {Moran}, J.~M., {Greenhill}, L.~J., \&
  {Argon}, A.~L. 2013, \apj, 775, 13, \dodoi{10.1088/0004-637X/775/1/13}

\bibitem[{{Hunter}(2007)}]{Hunter:2007}
{Hunter}, J.~D. 2007, Computing in Science and Engineering, 9, 90,
  \dodoi{10.1109/MCSE.2007.55}

\bibitem[{{Ibata} {et~al.}(2014){Ibata}, {Lewis}, {McConnachie}, {Martin},
  {Irwin}, {Ferguson}, {Babul}, {Bernard}, {Chapman}, {Collins}, {Fardal},
  {Mackey}, {Navarro}, {Pe{\~n}arrubia}, {Rich}, {Tanvir}, \&
  {Widrow}}]{Ibata2014}
{Ibata}, R.~A., {Lewis}, G.~F., {McConnachie}, A.~W., {et~al.} 2014, \apj, 780,
  128, \dodoi{10.1088/0004-637X/780/2/128}

\bibitem[{{Jacobs} {et~al.}(2009){Jacobs}, {Rizzi}, {Tully}, {Shaya},
  {Makarov}, \& {Makarova}}]{Jacobs2009}
{Jacobs}, B.~A., {Rizzi}, L., {Tully}, R.~B., {et~al.} 2009, \aj, 138, 332,
  \dodoi{10.1088/0004-6256/138/2/332}

\bibitem[{{Jang} \& {Lee}(2017)}]{Jang2017}
{Jang}, I.~S., \& {Lee}, M.~G. 2017, \apj, 835, 28,
  \dodoi{10.3847/1538-4357/835/1/28}

\bibitem[{{Joye} \& {Mandel}(2003)}]{Joye2003}
{Joye}, W.~A., \& {Mandel}, E. 2003, in Astronomical Society of the Pacific
  Conference Series, Vol. 295, Astronomical Data Analysis Software and Systems
  XII, ed. H.~E. {Payne}, R.~I. {Jedrzejewski}, \& R.~N. {Hook}, 489

\bibitem[{{Lee} {et~al.}(1993){Lee}, {Freedman}, \& {Madore}}]{Lee1993}
{Lee}, M.~G., {Freedman}, W.~L., \& {Madore}, B.~F. 1993, \apj, 417, 553,
  \dodoi{10.1086/173334}

\bibitem[{{Madore} \& {Freedman}(1995)}]{Madore1995}
{Madore}, B.~F., \& {Freedman}, W.~L. 1995, \aj, 109, 1645,
  \dodoi{10.1086/117391}

\bibitem[{{Madore} \& {Freedman}(2020)}]{Madore2020}
{Madore}, B.~F., \& {Freedman}, W.~L. 2020, \aj, 160, 170,
  \dodoi{10.3847/1538-3881/abab9a}

\bibitem[{{Madore} {et~al.}(2023{\natexlab{a}}){Madore}, {Freedman}, \&
  {Owens}}]{Madore2023b}
{Madore}, B.~F., {Freedman}, W.~L., \& {Owens}, K. 2023{\natexlab{a}}, \aj,
  166, 224, \dodoi{10.3847/1538-3881/ad022c}

\bibitem[{{Madore} {et~al.}(2023{\natexlab{b}}){Madore}, {Freedman}, {Owens},
  \& {Jang}}]{Madore2023}
{Madore}, B.~F., {Freedman}, W.~L., {Owens}, K.~A., \& {Jang}, I.~S.
  2023{\natexlab{b}}, \aj, 166, 2, \dodoi{10.3847/1538-3881/acd3f3}

\bibitem[{{Madore} {et~al.}(1997){Madore}, {Freedman}, \& {Sakai}}]{Madore1997}
{Madore}, B.~F., {Freedman}, W.~L., \& {Sakai}, S. 1997, in The Extragalactic
  Distance Scale, ed. M.~{Livio}, M.~{Donahue}, \& N.~{Panagia}, 239

\bibitem[{{Madore} {et~al.}(2009){Madore}, {Mager}, \& {Freedman}}]{Madore2009}
{Madore}, B.~F., {Mager}, V., \& {Freedman}, W.~L. 2009, \apj, 690, 389,
  \dodoi{10.1088/0004-637X/690/1/389}

\bibitem[{{Madore} {et~al.}(2018){Madore}, {Freedman}, {Hatt}, {Hoyt},
  {Monson}, {Beaton}, {Rich}, {Jang}, {Lee}, {Scowcroft}, \&
  {Seibert}}]{Madore2018}
{Madore}, B.~F., {Freedman}, W.~L., {Hatt}, D., {et~al.} 2018, \apj, 858, 11,
  \dodoi{10.3847/1538-4357/aab7f4}

\bibitem[{{Mager} {et~al.}(2008){Mager}, {Madore}, \& {Freedman}}]{Mager2008}
{Mager}, V.~A., {Madore}, B.~F., \& {Freedman}, W.~L. 2008, \apj, 689, 721,
  \dodoi{10.1086/592563}

\bibitem[{{Makarov} {et~al.}(2006){Makarov}, {Makarova}, {Rizzi}, {Tully},
  {Dolphin}, {Sakai}, \& {Shaya}}]{Makarov2006}
{Makarov}, D., {Makarova}, L., {Rizzi}, L., {et~al.} 2006, \aj, 132, 2729,
  \dodoi{10.1086/508925}

\bibitem[{{Marigo} {et~al.}(2017){Marigo}, {Girardi}, {Bressan}, {Rosenfield},
  {Aringer}, {Chen}, {Dussin}, {Nanni}, {Pastorelli}, {Rodrigues}, {Trabucchi},
  {Bladh}, {Dalcanton}, {Groenewegen}, {Montalb{\'a}n}, \& {Wood}}]{Marigo2017}
{Marigo}, P., {Girardi}, L., {Bressan}, A., {et~al.} 2017, \apj, 835, 77,
  \dodoi{10.3847/1538-4357/835/1/77}

\bibitem[{{McConnachie}(2012)}]{McConnachie2012}
{McConnachie}, A.~W. 2012, \aj, 144, 4, \dodoi{10.1088/0004-6256/144/1/4}

\bibitem[{{McQuinn} {et~al.}(2019){McQuinn}, {Boyer}, {Skillman}, \&
  {Dolphin}}]{McQuinn2019}
{McQuinn}, K. B.~W., {Boyer}, M., {Skillman}, E.~D., \& {Dolphin}, A.~E. 2019,
  \apj, 880, 63, \dodoi{10.3847/1538-4357/ab2627}

\bibitem[{{McQuinn} {et~al.}(2017){McQuinn}, {Skillman}, {Dolphin}, {Berg}, \&
  {Kennicutt}}]{McQuinn2017}
{McQuinn}, K. B.~W., {Skillman}, E.~D., {Dolphin}, A.~E., {Berg}, D., \&
  {Kennicutt}, R. 2017, \aj, 154, 51, \dodoi{10.3847/1538-3881/aa7aad}

\bibitem[{{M{\'e}ndez} {et~al.}(2002){M{\'e}ndez}, {Davis}, {Moustakas},
  {Newman}, {Madore}, \& {Freedman}}]{Mendez2002}
{M{\'e}ndez}, B., {Davis}, M., {Moustakas}, J., {et~al.} 2002, \aj, 124, 213,
  \dodoi{10.1086/341168}

\bibitem[{{Monachesi} {et~al.}(2016){Monachesi}, {Bell}, {Radburn-Smith},
  {Bailin}, {de Jong}, {Holwerda}, {Streich}, \& {Silverstein}}]{Monachesi2016}
{Monachesi}, A., {Bell}, E.~F., {Radburn-Smith}, D.~J., {et~al.} 2016, \mnras,
  457, 1419, \dodoi{10.1093/mnras/stv2987}

\bibitem[{{Monson} {et~al.}(2017){Monson}, {Beaton}, {Scowcroft}, {Freedman},
  {Madore}, {Rich}, {Seibert}, {Kollmeier}, \& {Clementini}}]{Monson2017}
{Monson}, A.~J., {Beaton}, R.~L., {Scowcroft}, V., {et~al.} 2017, \aj, 153, 96,
  \dodoi{10.3847/1538-3881/153/3/96}

\bibitem[{{Montalto} {et~al.}(2009){Montalto}, {Seitz}, {Riffeser}, {Hopp},
  {Lee}, \& {Sch{\"o}nrich}}]{Montalto2009}
{Montalto}, M., {Seitz}, S., {Riffeser}, A., {et~al.} 2009, \aap, 507, 283,
  \dodoi{10.1051/0004-6361/200912179}

\bibitem[{{Mould} \& {Kristian}(1986)}]{Mould1986}
{Mould}, J., \& {Kristian}, J. 1986, \apj, 305, 591, \dodoi{10.1086/164273}

\bibitem[{{Persson} {et~al.}(2004){Persson}, {Madore}, {Krzemi{\'n}ski},
  {Freedman}, {Roth}, \& {Murphy}}]{Persson2004}
{Persson}, S.~E., {Madore}, B.~F., {Krzemi{\'n}ski}, W., {et~al.} 2004, \aj,
  128, 2239, \dodoi{10.1086/424934}

\bibitem[{{Pietrinferni} {et~al.}(2004){Pietrinferni}, {Cassisi}, {Salaris}, \&
  {Castelli}}]{Pietrinferni2004}
{Pietrinferni}, A., {Cassisi}, S., {Salaris}, M., \& {Castelli}, F. 2004, \apj,
  612, 168, \dodoi{10.1086/422498}

\bibitem[{{Pietrinferni} {et~al.}(2006){Pietrinferni}, {Cassisi}, {Salaris}, \&
  {Castelli}}]{Pietrinferni2006}
{Pietrinferni}, A., {Cassisi}, S., {Salaris}, M., \& {Castelli}, F. 2006, \apj,
  642, 797, \dodoi{10.1086/501344}

\bibitem[{{Pietrzy{\'n}ski} {et~al.}(2008){Pietrzy{\'n}ski}, {Gieren},
  {Szewczyk}, {Walker}, {Rizzi}, {Bresolin}, {Kudritzki}, {Nalewajko}, {Storm},
  {Dall'Ora}, \& {Ivanov}}]{Pietrzynksi2008}
{Pietrzy{\'n}ski}, G., {Gieren}, W., {Szewczyk}, O., {et~al.} 2008, \aj, 135,
  1993, \dodoi{10.1088/0004-6256/135/6/1993}

\bibitem[{{Pietrzy{\'n}ski} {et~al.}(2013){Pietrzy{\'n}ski}, {Graczyk},
  {Gieren}, {Thompson}, {Pilecki}, {Udalski}, {Soszy{\'n}ski}, {Koz{\l}owski},
  {Konorski}, {Suchomska}, {Bono}, {Moroni}, {Villanova}, {Nardetto},
  {Bresolin}, {Kudritzki}, {Storm}, {Gallenne}, {Smolec}, {Minniti}, {Kubiak},
  {Szyma{\'n}ski}, {Poleski}, {Wyrzykowski}, {Ulaczyk}, {Pietrukowicz},
  {G{\'o}rski}, \& {Karczmarek}}]{Pietrzynski2013}
{Pietrzy{\'n}ski}, G., {Graczyk}, D., {Gieren}, W., {et~al.} 2013, \nat, 495,
  76, \dodoi{10.1038/nature11878}

\bibitem[{{Pietrzy{\'n}ski} {et~al.}(2019){Pietrzy{\'n}ski}, {Graczyk},
  {Gallenne}, {Gieren}, {Thompson}, {Pilecki}, {Karczmarek}, {G{\'o}rski},
  {Suchomska}, {Taormina}, {Zgirski}, {Wielg{\'o}rski}, {Ko{\l}aczkowski},
  {Konorski}, {Villanova}, {Nardetto}, {Kervella}, {Bresolin}, {Kudritzki},
  {Storm}, {Smolec}, \& {Narloch}}]{Pietrzynski2019}
{Pietrzy{\'n}ski}, G., {Graczyk}, D., {Gallenne}, A., {et~al.} 2019, \nat, 567,
  200, \dodoi{10.1038/s41586-019-0999-4}

\bibitem[{{Radburn-Smith} {et~al.}(2011){Radburn-Smith}, {de Jong}, {Seth},
  {Bailin}, {Bell}, {Brown}, {Bullock}, {Courteau}, {Dalcanton}, {Ferguson},
  {Goudfrooij}, {Holfeltz}, {Holwerda}, {Purcell}, {Sick}, {Streich}, {Vlajic},
  \& {Zucker}}]{Radburn-Smith2011}
{Radburn-Smith}, D.~J., {de Jong}, R.~S., {Seth}, A.~C., {et~al.} 2011, \apjs,
  195, 18, \dodoi{10.1088/0067-0049/195/2/18}

\bibitem[{{Reid} {et~al.}(2019){Reid}, {Pesce}, \& {Riess}}]{Reid2019}
{Reid}, M.~J., {Pesce}, D.~W., \& {Riess}, A.~G. 2019, \apjl, 886, L27,
  \dodoi{10.3847/2041-8213/ab552d}

\bibitem[{{Riess} {et~al.}(2016){Riess}, {Macri}, {Hoffmann}, {Scolnic},
  {Casertano}, {Filippenko}, {Tucker}, {Reid}, {Jones}, {Silverman},
  {Chornock}, {Challis}, {Yuan}, {Brown}, \& {Foley}}]{Riess2016}
{Riess}, A.~G., {Macri}, L.~M., {Hoffmann}, S.~L., {et~al.} 2016, \apj, 826,
  56, \dodoi{10.3847/0004-637X/826/1/56}

\bibitem[{{Riess} {et~al.}(2022){Riess}, {Yuan}, {Macri}, {Scolnic}, {Brout},
  {Casertano}, {Jones}, {Murakami}, {Anand}, {Breuval}, {Brink}, {Filippenko},
  {Hoffmann}, {Jha}, {D'arcy Kenworthy}, {Mackenty}, {Stahl}, \&
  {Zheng}}]{Riess2022}
{Riess}, A.~G., {Yuan}, W., {Macri}, L.~M., {et~al.} 2022, \apjl, 934, L7,
  \dodoi{10.3847/2041-8213/ac5c5b}

\bibitem[{{Rizzi} {et~al.}(2007){Rizzi}, {Tully}, {Makarov}, {Makarova},
  {Dolphin}, {Sakai}, \& {Shaya}}]{Rizzi2007}
{Rizzi}, L., {Tully}, R.~B., {Makarov}, D., {et~al.} 2007, \apj, 661, 815,
  \dodoi{10.1086/516566}

\bibitem[{{Sakai} {et~al.}(1996){Sakai}, {Madore}, \& {Freedman}}]{Sakai1996}
{Sakai}, S., {Madore}, B.~F., \& {Freedman}, W.~L. 1996, \apj, 461, 713,
  \dodoi{10.1086/177096}

\bibitem[{{Sakai} {et~al.}(1997){Sakai}, {Madore}, {Freedman}, {Lauer},
  {Ajhar}, \& {Baum}}]{Sakai1997}
{Sakai}, S., {Madore}, B.~F., {Freedman}, W.~L., {et~al.} 1997, \apj, 478, 49,
  \dodoi{10.1086/303768}

\bibitem[{{Salaris} \& {Cassisi}(1997)}]{Salaris1997}
{Salaris}, M., \& {Cassisi}, S. 1997, \mnras, 289, 406,
  \dodoi{10.1093/mnras/289.2.406}

\bibitem[{{Salaris} {et~al.}(2002){Salaris}, {Cassisi}, \&
  {Weiss}}]{Salaris2002}
{Salaris}, M., {Cassisi}, S., \& {Weiss}, A. 2002, \pasp, 114, 375,
  \dodoi{10.1086/342498}

\bibitem[{{Salaris} \& {Girardi}(2005)}]{Salaris2005}
{Salaris}, M., \& {Girardi}, L. 2005, \mnras, 357, 669,
  \dodoi{10.1111/j.1365-2966.2005.08689.x}

\bibitem[{{Sandage}(2006)}]{Sandage2006}
{Sandage}, A. 2006, \aj, 131, 1750, \dodoi{10.1086/500012}

\bibitem[{{Schlafly} \& {Finkbeiner}(2011)}]{Schlafly2011}
{Schlafly}, E.~F., \& {Finkbeiner}, D.~P. 2011, \apj, 737, 103,
  \dodoi{10.1088/0004-637X/737/2/103}

\bibitem[{{Schlegel} {et~al.}(1998){Schlegel}, {Finkbeiner}, \&
  {Davis}}]{Schlegel1998}
{Schlegel}, D.~J., {Finkbeiner}, D.~P., \& {Davis}, M. 1998, \apj, 500, 525,
  \dodoi{10.1086/305772}

\bibitem[{{Serenelli} {et~al.}(2017){Serenelli}, {Weiss}, {Cassisi}, {Salaris},
  \& {Pietrinferni}}]{Serenelli2017}
{Serenelli}, A., {Weiss}, A., {Cassisi}, S., {Salaris}, M., \& {Pietrinferni},
  A. 2017, \aap, 606, A33, \dodoi{10.1051/0004-6361/201731004}

\bibitem[{{Tully} {et~al.}(2009){Tully}, {Rizzi}, {Shaya}, {Courtois},
  {Makarov}, \& {Jacobs}}]{Tully2009}
{Tully}, R.~B., {Rizzi}, L., {Shaya}, E.~J., {et~al.} 2009, \aj, 138, 323,
  \dodoi{10.1088/0004-6256/138/2/323}

\bibitem[{{Valenti} {et~al.}(2004){Valenti}, {Ferraro}, \&
  {Origlia}}]{Valenti2004}
{Valenti}, E., {Ferraro}, F.~R., \& {Origlia}, L. 2004, \mnras, 354, 815,
  \dodoi{10.1111/j.1365-2966.2004.08249.x}

\bibitem[{{Virtanen} {et~al.}(2020){Virtanen}, {Gommers}, {Oliphant},
  {Haberland}, {Reddy}, {Cournapeau}, {Burovski}, {Peterson}, {Weckesser},
  {Bright}, {van der Walt}, {Brett}, {Wilson}, {Millman}, {Mayorov}, {Nelson},
  {Jones}, {Kern}, {Larson}, {Carey}, {Polat}, {Feng}, {Moore}, {VanderPlas},
  {Laxalde}, {Perktold}, {Cimrman}, {Henriksen}, {Quintero}, {Harris},
  {Archibald}, {Ribeiro}, {Pedregosa}, {van Mulbregt}, \& {SciPy 1. 0
  Contributors}}]{2020SciPy-NMeth}
{Virtanen}, P., {Gommers}, R., {Oliphant}, T.~E., {et~al.} 2020, Nature
  Methods, 17, 261, \dodoi{10.1038/s41592-019-0686-2}

\bibitem[{{Weiss} \& {Schlattl}(2008)}]{Weiss2008}
{Weiss}, A., \& {Schlattl}, H. 2008, \apss, 316, 99,
  \dodoi{10.1007/s10509-007-9606-5}

\bibitem[{{Williams} {et~al.}(2014){Williams}, {Lang}, {Dalcanton}, {Dolphin},
  {Weisz}, {Bell}, {Bianchi}, {Byler}, {Gilbert}, {Girardi}, {Gordon},
  {Gregersen}, {Johnson}, {Kalirai}, {Lauer}, {Monachesi}, {Rosenfield},
  {Seth}, \& {Skillman}}]{Williams2014}
{Williams}, B.~F., {Lang}, D., {Dalcanton}, J.~J., {et~al.} 2014, \apjs, 215,
  9, \dodoi{10.1088/0067-0049/215/1/9}

\bibitem[{{Williams} {et~al.}(2021){Williams}, {Durbin}, {Dalcanton}, {Lang},
  {Girardi}, {Smercina}, {Dolphin}, {Weisz}, {Choi}, {Bell}, {Rosolowsky},
  {Skillman}, {Koch}, {Lindberg}, {Hagen}, {Gordon}, {Seth}, {Gilbert},
  {Guhathakurta}, {Lauer}, \& {Bianchi}}]{Williams2021}
{Williams}, B.~F., {Durbin}, M.~J., {Dalcanton}, J.~J., {et~al.} 2021, \apjs,
  253, 53, \dodoi{10.3847/1538-4365/abdf4e}

\bibitem[{{Wu} {et~al.}(2014){Wu}, {Tully}, {Rizzi}, {Dolphin}, {Jacobs}, \&
  {Karachentsev}}]{Wu2014}
{Wu}, P.-F., {Tully}, R.~B., {Rizzi}, L., {et~al.} 2014, \aj, 148, 7,
  \dodoi{10.1088/0004-6256/148/1/7}

\bibitem[{{Yuan} {et~al.}(2019){Yuan}, {Riess}, {Macri}, {Casertano}, \&
  {Scolnic}}]{Yuan2019}
{Yuan}, W., {Riess}, A.~G., {Macri}, L.~M., {Casertano}, S., \& {Scolnic},
  D.~M. 2019, \apj, 886, 61, \dodoi{10.3847/1538-4357/ab4bc9}

\end{thebibliography}

\end{document}